\documentclass{raa_twocolumn}            

\usepackage{graphicx,times}             
\usepackage{natbib}
\usepackage{amssymb,amsmath}
\usepackage{lscape}
\usepackage{multirow}
\bibpunct{(}{)}{;}{a}{}{,}

\usepackage[a4paper=true, pagebackref=true]{hyperref}
\hypersetup{colorlinks = true, linkcolor = green, anchorcolor = red, citecolor = blue, filecolor = red, pagecolor = red, urlcolor = red}

\usepackage{multirow}
\usepackage{ulem}
\usepackage{color}
\usepackage{booktabs}
\usepackage{hyperref}
\usepackage{longtable}
\usepackage{amssymb,mathrsfs,amsmath}
\usepackage{supertabular}

\begin{document}

   \title{Power spectral properties of the soft spectral states in four black hole transients}
   
   \volnopage{{\bf 20XX} Vol.\ {\bf X} No. {\bf XX}, 000--000}      
   \setcounter{page}{1}          

   \author{Dong-Ming Mao
      \inst{1, 2}
   \and Wen-Fei Yu
      \inst{1}\thanks{Corresponding author}
          }

   \institute{Shanghai Astronomical Observatory, Chinese Academy of Sciences,
             Shanghai 200030, China; \hs\hs\hs {\it wenfei@shao.ac.cn}\\
   \and
            University of Chinese Academy of Sciences, Beijing 100049, China\\
\vs\no
   {\small Received~~2021.1.17 accepted~~2021.2.28}}

\abstract{The X-ray variability in the soft X-ray spectral state of black hole binaries is primarily characterized by a power-law noise (PLN), which is thought to originate from the propagation of the modulation in the mass accretion rate of a standard accretion disk flow. Such a PLN has also been revealed in the disk spectral component in the hard and the intermediate states in several black hole binaries. Here we present an investigation of the {\it Rossi} X-ray Timing Explorer (RXTE) observations of four black hole transients in which soft spectral states were observed twenty times or more. We show that in the soft spectral state, the PLN index varied in a large range between -1.64 and -0.62, and the fractional rms variability calculated in the 0.01 -- 20 Hz frequency range reached as large as 7.67\% and as low as 0.83\%. Remarkably, we have found the evidence of an inclination dependence of the maximal fractional rms variability, the averaged fractional rms variability and the fractional rms variability of the median in the sample based on current knowledge of inclination of black hole binaries. An inclination dependence has only been predicted in early magnetohydrodynamic simulations of isothermal disks limited to a high-frequency regime. In theory, the noise index is related to the physics of inward propagation of disk fluctuations, while the fractional rms amplitude reflects the intrinsic properties of the magnetohydrodynamic nature of the accretion flow. Our results therefore suggest that X-ray variability in the soft state can be used to put constraints on the properties of the accretion flow as well as the inclination of the accretion disk.
}

   \authorrunning{D.-M. Mao, W.-F. Yu}            
   \titlerunning{Black hole soft state}  

   \maketitle

\section{Introduction}
Black hole X-ray binaries (BHXBs) are binary systems that consist of a black hole as the primary while the companion star feeds mass through either a Roche lobe overflow or a stellar wind \citep{1976A&A....49..327L, 2002apa..book.....F}. In most cases, due to the transfer of the angular momentum outwardly, the mass flow will form an optically thick and geometrically thin accretion disk around the BH \citep{1973A&A....24..337S}, or certain alternatives in different regimes, for example the radiatively inefficient accretion flow \citep[see][]{1997ApJ...489..865E}. The gravitational potential energy of the accreted matter dissipated in the innermost accretion flow would primarily contribute to the X-ray emission, making them one of the major populations of the X-ray sources in our Galaxy. BHXBs have been observed as both persistent and transient X-ray sources. Cygnus X$-$1 is one of the classical persistent BHXBs, which was discovered in the 1960s \citep{1965Sci...147..394B}. Most other BHXBs in our Galaxy belongs to the transient category, such as A0620$-$00 \citep{1975Natur.257..656E}. As we have learned, these black hole transients usually spend most of their time in the so-called quiescent state ($L_{x} < \sim {10}^{33} {\rm erg\ s^{-1}}$).  Occasionally they turn into outbursts that would last days, weeks, years or even longer \citep{1997ApJ...491..312C, 2015ApJ...805...87Y} and go through different X-ray spectral states \citep{2006ARA&A..44...49R, 2009ApJ...701.1940Y}.

X-ray spectral state transitions in BHXBs was discovered about fifty years ago. \cite{1972ApJ...177L...5T} observed a spectral change in Cyg X$-$1 in which the hard X-ray flux (10--20 keV) increased and the soft X-ray flux (2--6 keV) decreased. A large number of X-ray observations to date have shown that most of the BHXBs show a similar spectral behavior. According to the observed energy spectra and power spectra, the behavior of BHXBs is characterized into a series of X-ray spectral states and transitions between them \citep[][and references therein]{1992ApJ...391L..21M, 1993ApJ...403L..39M, 1997ApJ...489..865E}. According to the pictures established by the RXTE timing and spectral observations, there are three major spectral states, namely, the soft state, the hard state and the intermediate state  (which includes both hard and soft intermediate states) \citep{2001ApJS..132..377H, 2010LNP...794...53B}. Transient BHXBs typically show a hard state during the early rising phase and the later decay phase of their outbursts. The corresponding X-ray broad-band energy spectrum is very hard (photon index, $\Gamma\sim1.4-2.1$) sometimes with a high energy cutoff around 50--100 keV, which is generally associated to the Comptonization of the disk soft photons by hot electrons in the corona \citep{1998MNRAS.298..729D, 2010LNP...794...17G}. During the rising phase or even at the flux peak or decay phase of an X-ray outburst, a transient BHXB might proceed with a transition into the soft spectral state, during which sometimes a transient BHXB would go through the intermediate X-ray states between the hard X-ray state and the soft X-ray state. In such a state, the X-ray energy spectrum is dominated by a thermal component which is well modeled by a multi-color blackbody from an accretion disk, with a temperature with a typical value of about 1 keV, representative of the temperature at the inner most disk edge \citep{1973A&A....24..337S, 1997ApJ...477L..95Z, 1999MNRAS.309..496G, 2010LNP...794...53B, 2016ASSL..440...61B}. Besides the differences in the X-ray spectral properties among the spectral states, BHXBs also show distinct properties in the Fourier power spectra in different spectral states as well, thus it is generally believed that both energy spectral and temporal spectral properties are needed to distinguish these black hole X-ray states. 

X-ray variability is a useful probe of the structure of the accretion flow and its evolution \citep{1999ApJ...514..939W, 2008ApJ...675.1407K}. To investigate black hole X-ray variability, conventionally people use Fourier power density spectra (PDS) and sometimes the fractional root mean square (rms) \citep{1983ApJ...266..160L, 1989ARA&A..27..517V, 2002ApJ...572..392B}. During the X-ray hard state, X-ray variability is very strong, with the fractional rms as high as 40\%, and the PDS is well described by the superposition of some band-limited noise (BLN) components and quasi-periodic oscillations. In the X-ray soft state, BHXBs lack of strong variability at higher frequencies. The PDS in the soft spectral state is primarily described by a power-law noise (PLN) \citep{1994ApJ...435..398M, 1997ApJ...474L..57C, 2006ARA&A..44...49R, 2010LNP...794...17G}. Among the BHXBs, Cyg X$-$1 is often considered as a classical example for comparison. In the hard spectral state, its power spectra are flat from low frequencies up to a certain critical frequency, at which they break into a power-law component with a $f^{-1}$ slope \citep{1990A&A...227L..33B, 1999ApJ...510..874N}. In the soft spectral state, its power spectra are characterized by a $f^{-1}$ component up to at least $\sim$ 10 Hz,  above which it might steepen as $f^{-2}$ \citep{1997ApJ...474L..57C, 2000MNRAS.316..923G}. However, since it is a wind-accretion HMXB system and has probably remained in a narrow range of the mass accretion rate, Cyg X$-$1 is probably not the best target to compare with black hole transients. One of the representative transient BHXBs is probably the black hole transient GX~339$-$4. Its fractional rms variability has been seen less than 5\% \citep {2011MNRAS.418.2292M,2011BASI...39..409B} and some times can be as low as 1\% in its soft spectral state \citep{2005A&A...440..207B,2010LNP...794...53B}. 

In the black hole soft spectral state, the domination of the standard disk component in the energy spectra and the obvious low fractional rms variability indicate a lack of variability in the emission of the optically thick accretion disk, as compared with the variability in the emission from the corona which contributes to the hard spectral component and dominates in the hard and intermediate states. The same conclusion can be drawn for the cold, standard disk in the hard state and the intermediate state, as demonstrated at least in several black hole binaries observed in the past. \cite{2013ApJ...770..135Y} studied the evolution of the power spectra and energy spectra of the transient BHXB MAXI J1659$-$152, which has a low Galactic hydrogen absorption due to its location. They found that the PLN is associated with a disk spectral component in both hard and intermediate states below 2 keV. This unifies our understanding of the variability properties of standard accretion disks across all the spectral states as of PLN form. To interpret the low rms variability and the PL shape of the X-ray PDS observed in the soft state, one of the mechanisms proposed is the viscosity fluctuations in the accretion disk \citep{1997MNRAS.292..679L}. An MHD simulation of propagating fluctuations in a geometrically thin and optically thick accretion disk around a black hole, given by \cite{2016ApJ...826...40H}, shows that the effective $\alpha$-parameter in the disk is highly variable due to the magneto-rotational instability (MRI) \citep{1991ApJ...376..214B} and the “disk dynamo”. The resulting power spectrum of their synthetic light curve shows a power-law shape, as what we have observed in the soft spectral state or in the soft X-ray band when it is dominated by the disk component in either the hard or the intermediate state. 

The X-ray variability in the black hole soft state is thus a crucial element for our understanding of the standard accretion disk flow in BHXBs and its magnetic field properties. For this reason, we made use of the Rossi X-ray Timing Explorer (RXTE) observations of four transient BHXBs, namely 4U~1630$-$47, H~1743$-$322, XTE~J1859$+$226 and XTE~J1817$-$330, in which twenty of RXTE observations of the likely soft spectral states of those black hole transients exist, in order to investigate the statistical properties of the PLN seen in the soft states. Specifically, we intend to investigate the extremes of the PL indices and the fractional rms variability for each individual sources, which allow the investigation of the physical regimes unrelated or related to the parameters of the black holes or binaries. 

\section{Observations}
The {\it Rossi} X-ray Timing Explorer (RXTE) was launched in late 1995 and had performed for about 16 years of observations. It can provide high time resolution observations in combination with the moderate spectral resolution and broadband coverage, capable of exploring the X-ray variability of Galactic black hole and neutron star binaries in different states. There had been about twenty BHXBs observed with the Proportional Counter Array (PCA) on board of the RXTE. We intend to study the properties of the soft spectral states observed in those black hole transients instead. We focus on four black hole transients of which at least twenty observations of the soft spectral states have been performed. 4U~1630$-$47 is a recurrent black hole X-ray transient with outbursts in every 600 to 690 days or so \citep{1997MNRAS.291...81K}. H~1743$-$322 is another black hole transient \citep{1977IAUC.3106....3K} with very frequent outbursts in the RXTE era \citep[e.g.][]{2009ApJ...698.1398M, 2010MNRAS.408.1796M}. It also had quite some outbursts that did not reach the soft spectral state \citep{2009MNRAS.398.1194C, 2015MNRAS.452.3666S}. The black hole transient XTE~J1859$+$226 in our sample is a typical black hole transient as well \citep{2002MNRAS.331..169H}. One more source in our sample is the black hole transient XTE~J1817$-$330, which was discovered on 2006 January 26 \citep{2006ATel..714....1R}. The PCA pointed observations of these black hole transients performed by the RXTE allow us to perform a statistical study of the Fourier power spectra corresponding to the soft X-ray spectral state, with the number of observations of each target ranging from several tens to even more than a hundred RXTE observations (see Figure \ref{ratio} and Table \ref{minmax}, \ref{minmax2} and \ref{fitting result}). The averaged properties of the power spectra and the intrinsic differences among the power spectra of individual sources thus can be investigated. 

\section{Sample selection}
\begin{figure*}
\centering
\includegraphics[width=0.85\textwidth]{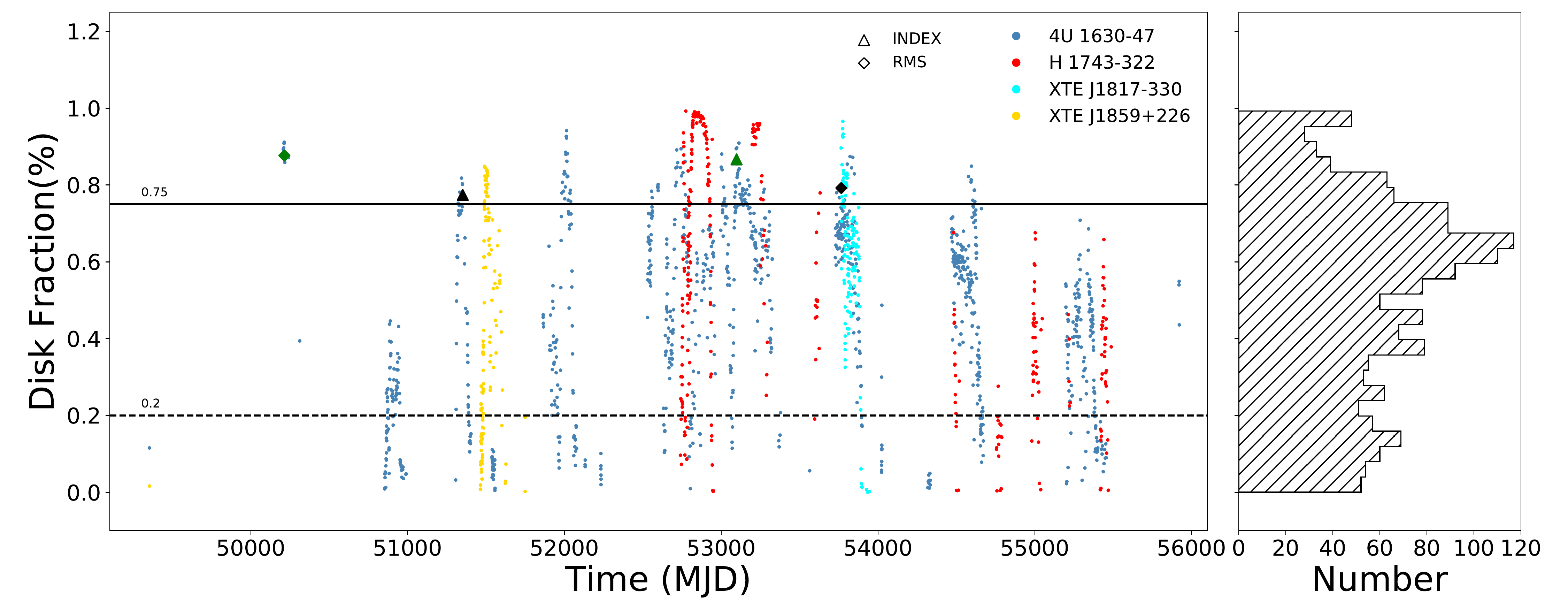}
\caption{The corresponding fraction of the energy flux of the disk component is relative to the total energy flux in the RXTE/PCA observations of 4U~1630$-$47 (blue), H~1743$-$322 (red), XTE~J1817$-$330 (cyan) and XTE~J1859$+$226 (yellow), respectively. In the left panel, the solid line at 0.75 and the dashed line at 0.2 indicate the disk fraction criteria used for the soft state and the hard state, respectively. As an initial approach, we selected the sample of the observations corresponding to the soft spectral state for all the four sources based on disk fraction alone. The triangles and the diamonds mark the observations of which the power spectra are shown in Figure \ref{index} and Figure \ref{rms} with the same colors. In the right panel, we show the distribution of the number of observations in relation to the disk fraction. The number of observations of which disk fraction is above the criteria of 0.75 is 124 for 4U~1630$-$47, 106 for H~1743$-$322, 31 for XTE~J1817$-$330 and 24 for XTE~J1859$+$226, respectively.}
\label{ratio}
\end{figure*}

Since a black hole soft state is determined by both energy spectral and power spectral properties \citep{2001ApJS..132..377H}, we used a threshold of 0.75 for the disk fraction between the flux of the disk component and the total energy flux \citep{2006ARA&A..44...49R}, and with the additional requirement of a power-law only power spectrum to select a PLN-only sample, and with a power-law dominated power spectrum to select a PLN-dominated sample in those observations corresponding to the soft spectral state. Figure \ref{ratio} shows the disk fraction of the RXTE observations of the four BHXBs in the period between MJD 49900 and MJD 55600. The histogram in the right panel of Figure \ref{ratio} shows the distribution of the number of the observations of the four sources.

\begin{figure*}
\centering
 \includegraphics[width=0.6\textwidth]{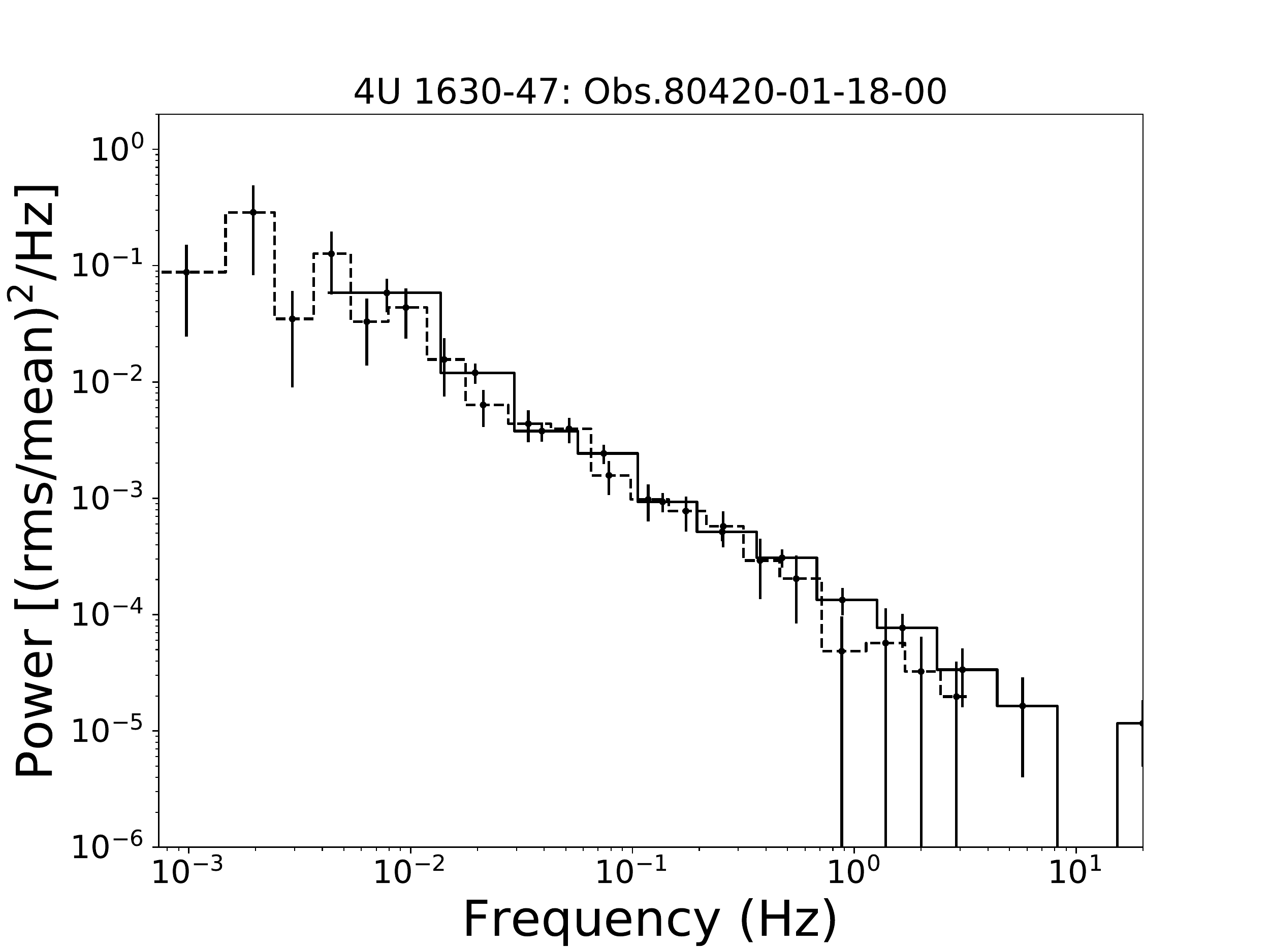}
\caption{An example of the power spectra corresponding to an observation (Observation ID 80420-01-18-00) of 4U~1630$-$472 when the source was in the soft state. The solid and dashed lines represent the power spectra calculated from the original PCA light curve and the net light curve from the PCA standard products, respectively.}
\label{long}
\end{figure*}

By including the power spectra generated from the net light curves obtained from the PCA standard products, we obtained the X-ray power spectra covering the frequency range as low as $\rm \sim 10^{-3}$ Hz and as high as 2048 Hz (Figure \ref{long}). We found that the power spectra obtained from the standard products and the high time resolution data are consistent with each other. So based on the power spectra obtained from the high time resolution data alone, we are able to address the properties of the PLN extending to low frequencies, as the power spectra do not suffer from significant particle background modulation.  

\subsection{Energy spectral and power spectral analysis}
We first made use of the RXTE/PCA standard products to identify those observations in which our targets were in the soft energy spectral states. We applied a spectral model $wabs \cdot (diskbb + powerlaw)$ in XSPEC to fit the energy spectra in the energy band 3--20 keV. We calculated the fluxes of the $diskbb$ component and the $powerlaw$ component, respectively. The ratio between the flux of the $diskbb$ component and the total energy flux is shown in Figure \ref{ratio} for all the observations of all the four targets. For the purpose of our study, we took a disk-flux fraction threshold of 75\% in the 3--20 keV band, which is more rigorous than that used as the spectral threshold in a wider and softer energy range 2--20 keV in the definition of the thermal-dominated state or the soft state discussed in \cite{2006ARA&A..44...49R}. 

In order to study rapid X-ray variability, we made use of high time resolution data (mostly with a time resolution of 250 $\mu$s or better) in each of the RXTE observations, i.e., data in Single Bit mode, Bin mode or Event mode, and extracted photon arrival time series from the original PCA data and then combined them into a time series with a time resolution of $\sim$ 488.28 $\mu$s (as multiples of the original time stamps). Then we calculated the corresponding Fourier power spectra for every 128 s data segment and then obtained the corresponding average power spectra in the frequency range up to 2048 Hz. The white noise level could be simply estimated as a constant at the averaged power in the frequency range above 1800 Hz. However, since the Poisson level is influenced by the instrument dead time which varies with frequency \citep{1995ApJ...449..930Z, 2006ApJS..163..401J}, we applied a shift of the Zhang-model to fit the power spectra above 1800 Hz, following the previous practice described in \citep{2004PhDT.......415K}.

The dead-time effect in the PCA data is described in \cite{1995ApJ...449..930Z} as:
\begin{equation}
\begin{split}
    P_{v} = 2-4 r_{0} t_{d}\left(1-\frac{t_{d}}{2 t_{b}}\right)-2\left(\frac{N-1}{N}\right) r_{0} t_{d} \times \\
\left(\frac{t_{d}}{t_{b}}\right) \cos \left(2 \pi v t_{b}\right)+2 r_{\rm vle} t_{\rm vle}^{2}\left(\frac{\sin \pi t_{\rm vle} v}{\pi t_{\rm vle} v}\right)^{2}
\end{split}
\end{equation}
Here, $P_{v}$ is the Poisson level at frequency $v$, $N$ is the number of frequency points, $r_{0}$ is the count rate per proportional counter unit (PCU), $r_{\rm vle}$ is the rate of very large events (VLE) per PCU, $t_{d}$ and $t_{\rm vle}$ is the dead time for event and VLE, respectively, $t_{b}$ is the bin size. Following previous approaches,  we took $r_{vle}$ = 200 $\rm counts\ s^{-1}\ PCU^{-1}$ \citep{2012AJ....143..148P}, $t_{d}$ = 10 $\mu$s  \citep{2006ApJS..163..401J}, and $t_{\rm vle}$ = 170 $\mu$s \citep{2006ApJS..163..401J} in our investigation.

\subsection{The PLN-only sample}
Conventionally a black hole soft state is determined by both the X-ray energy spectra and the power spectra. We therefore applied power spectral criteria to further select the sample of the observations corresponding to the soft state after an initial selection based on the X-ray energy spectra. For the observations in which the disk-flux fraction is above 75\%, we further fitted the corresponding rms-normalized power spectra in which the white noise level as described above was subtracted with a simple power-law model that characterizes the PLN component. We limited our model fits in the frequency range between 0.01 Hz and 20 Hz in which we calculated the integrated rms, due to limited statistics towards higher frequencies. In order to obtain good fits, we also re-binned the power spectra in an appropriate logarithmic factor, i.e., 1.4 for most of the power spectra (only in nine observations we are required to use a larger factor of 2 to obtain reasonable fits due to poor statistics). We took reduced $\chi^{2}$ less than 2.0 as the criteria to identify those observations corresponding to the PLN-only sample of the soft state. In the end, by applying the requirements that the disk-flux fraction is above 75\% and the X-ray power spectrum is statistically consistent with a single power-law, we determined the so-called PLN-only sample of black hole soft states.

\subsection{The PLN-dominated sample}
For those observations in which the disk-flux fraction is above 75\%, a band-limited noise (BLN) component might still exist in addition to the PLN. In order to determine the contribution of a potential BLN component in all these observations, we further used the model composed of a zero-centered Lorentzian representing for the BLN and a power-law component representing for the PLN to fit all the power spectra of the observations meeting the requirement of the energy spectral criteria of the soft spectral state alone. Subsequently, we took the criteria that the reduced $\chi^{2}$ less than 2.0 for a power-law plus a zero-centered Lorentzian model fit and the fractional rms of the PLN component $>80\%$ of the total rms to identify the sample of observations that the PLN component dominates. In several observations of the sample, the power spectra can be fitted fairly well with the single power-law model. All those observations form an independent sample of observations of the black hole soft state as compared with the PLN-only sample. We refer the sample as the PLN-dominated sample of the black hole soft state.

\section{Detailed Results}
\subsection{Characterizing the X-ray variability with power spectral analysis}
\subsubsection{Results of the PLN-only sample}
\begin{figure}
\centering
 \includegraphics[width=0.47\textwidth]{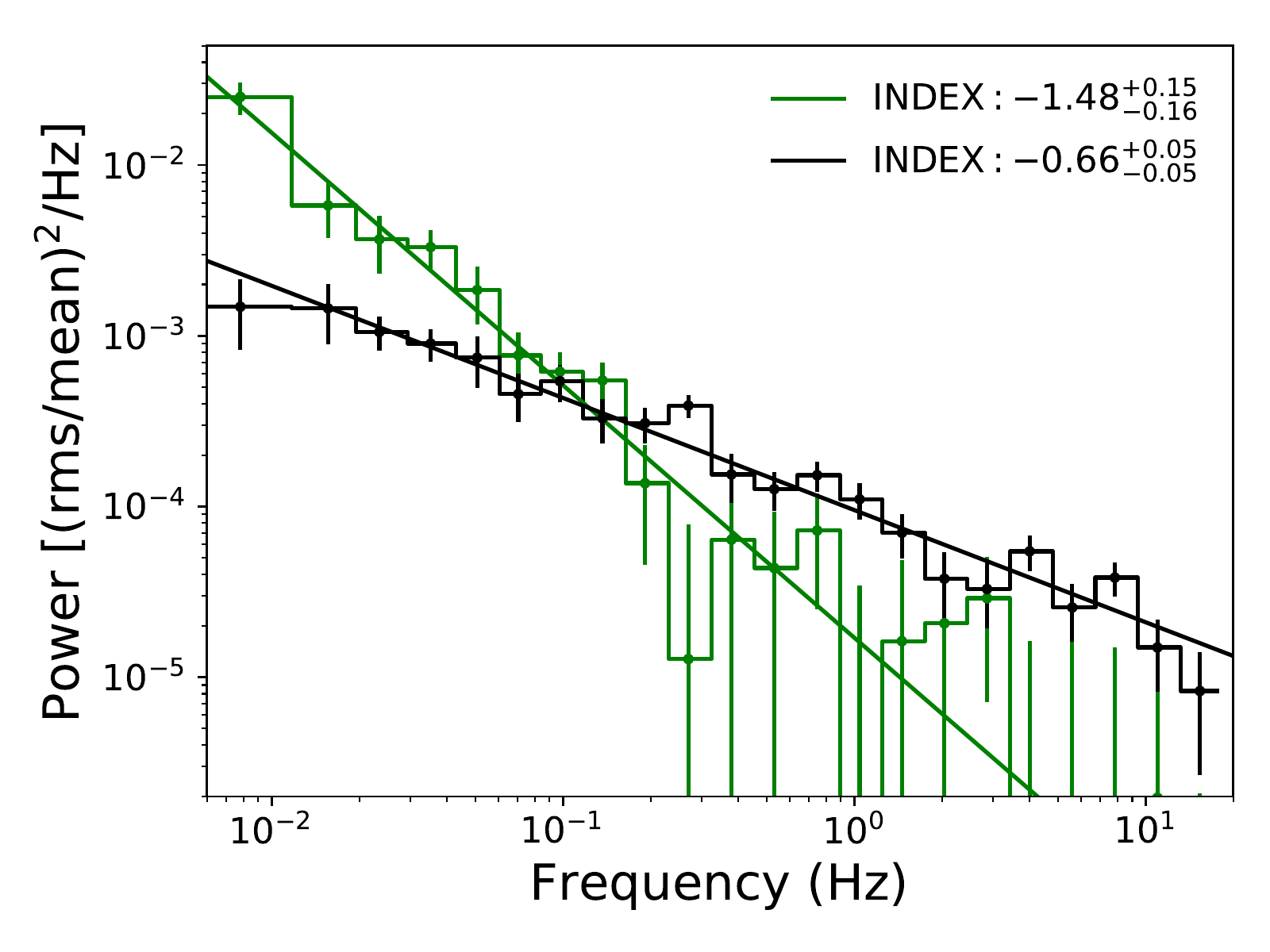}
\caption{Two representative RXTE/PCA power spectra showing distinct PLN indices.} 
\label{index}
\end{figure}

As noted above, the power spectra of the observation sample of the soft state can be fitted with a simple power-law (PL) model. The range of the fractional rms and the PL index for the sample we measured are listed in Table \ref{minmax}. We found that the PL indices are distributed around -0.97. The distribution of the PL index can be modeled with a Gaussian function to show its intrinsic spread. The minimum was around -1.4 and the maximum reached -0.7 and beyond. The exact PL index ranges from -1.48 to -0.66. We also show the power spectra corresponding to the observations in which the maximum and the minimum value of the PL index were seen in Figure \ref{index}. The list of the observation IDs, the disk fractions and the best-fit parameters are shown in Table \ref{fitting result}.

\begin{figure*}
\centering
 \includegraphics[width=0.85\textwidth]{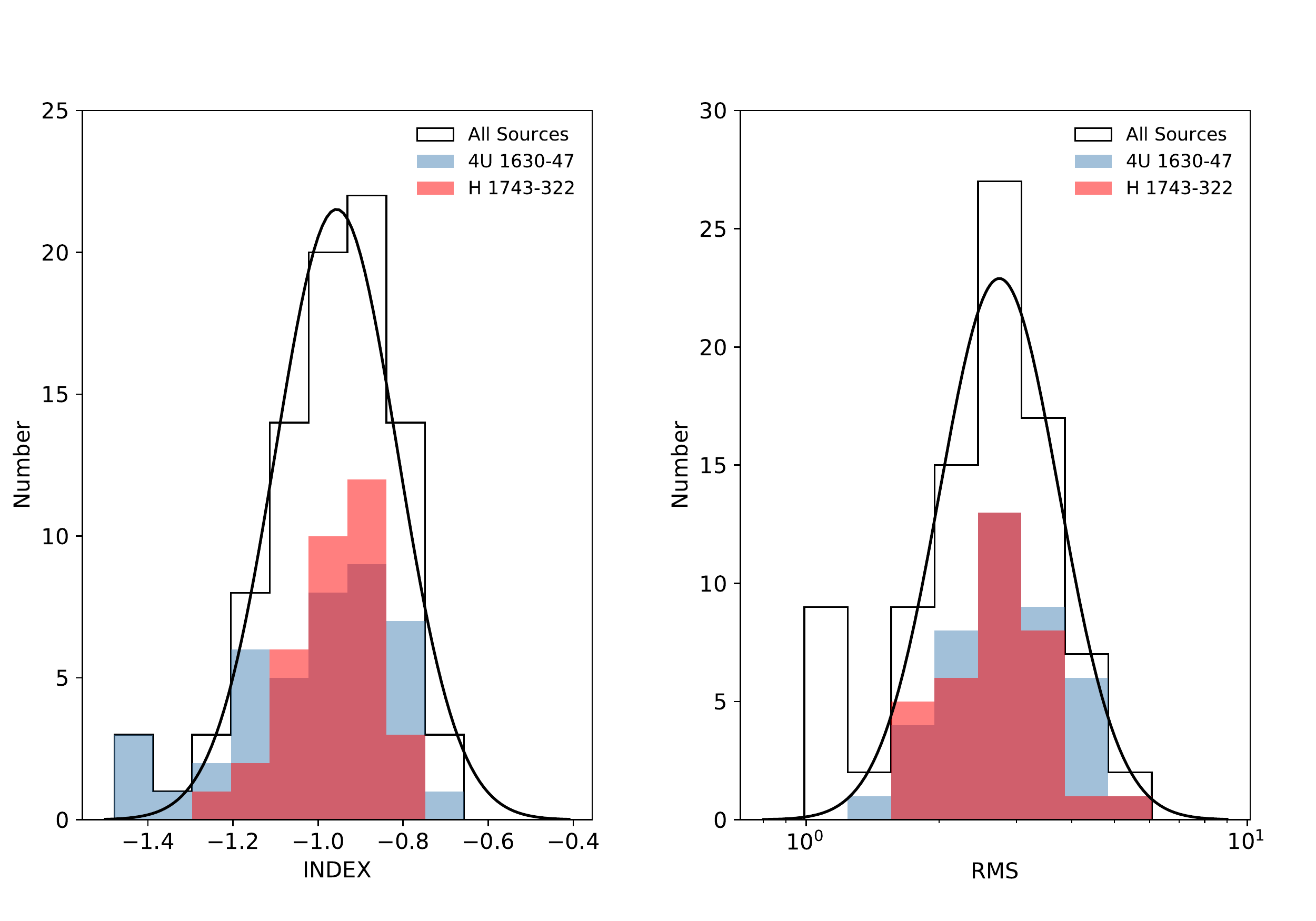}
\caption{The distributions of the PLN indices (left) and the fractional rms variability (right) obtained from the power spectra corresponding to the soft spectral state of the PLN-only sample. Both the distribution of the PLN indices and the distribution of the rms variability (in logarithmic scale) are over-plotted with the best-fit Gaussian functions to show their large spread. We also plot the distribution of the PL indices and fractional rms separately for 4U~1630$-$47 (blue) and H~1743$-$422 (red).}
\label{hist}
\end{figure*}

In order to investigate the statistical properties of the variability of these BHXBs, we calculated the fractional rms variability by integrating the PLN model component in the frequency range between 0.01 and 20 Hz from the best-fit model \citep{2013ApJ...770..135Y}. First, we zoomed in the observations with the rms variability significantly detected by more than $3\sigma$. This would help us put constraint on the maximum rms variability permitted by physics in the soft state. We obtained the observed range of the fractional rms of the PLN component, represented by the range between the minimal and the maximal fractional rms in the soft state. These measurements are listed in Table \ref{minmax}. The fractional rms of the PLN-only sample is on average around 2.7\% with the maxima of 6.07\% and the minima of 0.99\%. The distribution of the fractional rms of those observations is shown in the right panel of Figure \ref{hist}. The minima and maxima of the fractional rms corresponding to each of the four sources are also listed in Table \ref{minmax}. A dedicated investigation of those observations with the fractional rms less than 3$\sigma$ above zero was performed as well. The total number of these observations was 55. The averaged fractional rms is 1.78$\pm$1.56\%, larger than the minima obtained previously. Thus these observations likely lack of photon statistics instead of having lower fractional rms.

\begin{table*}
\centering
\caption{The fractional rms of the PLN component and the PL indices of the PLN-only sample}
\label{minmax}

\begin{tabular}{llcccc}
\toprule
          Source  &   Measurements  &              fractional rms (\%)  &         Obs ID &             index &         Obs ID \\
\midrule

\multirow{1}{*}{4U~1630$-$47} 
 & minimum & ${1.50}_{-1.11}^{+2.29}$ & 60118-01-05-00 & ${-1.48}_{-0.16}^{+0.15}$ & 40418-01-18-05 \\ [5pt]
 & maximum & ${5.53}_{-0.88}^{+0.91}$ & 70113-01-40-00 & ${-0.69}_{-0.29}^{+0.22}$ & 90128-01-14-00 \\ [5pt]
\\
\multirow{1}{*}{H~1743$-$322} 
 & minimum & ${1.73}_{-0.43}^{+0.49}$ & 90058-10-02-00 & ${-1.24}_{-0.20}^{+0.17}$ & 90421-01-03-00 \\ [5pt]
 & maximum & ${6.07}_{-0.20}^{+0.20}$ & 80144-01-03-00 & ${-0.77}_{-0.06}^{+0.06}$ & 80146-01-92-00 \\ [5pt]
\\
\multirow{1}{*}{XTE~J1817$-$330}
 & minimum & ${0.99}_{-0.30}^{+0.35}$ & 92082-01-01-01 & ${-1.03}_{-0.14}^{+0.13}$ & 91110-02-04-00 \\ [5pt]
 & maximum & ${2.70}_{-0.15}^{+0.15}$ & 91110-02-21-00 & ${-0.66}_{-0.05}^{+0.05}$ & 91110-02-21-00 \\ [5pt]
\\
\multirow{1}{*}{XTE~J1859$+$226}
 & minimum & ${1.02}_{-0.37}^{+0.46}$ & 40124-01-48-01 & ${-1.05}_{-1.41}^{+0.38}$ & 40124-01-53-01 \\ [5pt]
 & maximum & ${1.28}_{-0.60}^{+0.76}$ & 40124-01-53-02 & ${-0.77}_{-0.65}^{+0.32}$ & 40124-01-53-02 \\
\bottomrule
\\
\multicolumn{6}{l}{$Note.$ The uncertainties of the fractional rms and the PL indices correspond to the confidence level of 90\%}\\
\end{tabular}
\end{table*}

\begin{table*}
\centering
\caption{The fractional rms of PLN component and the PL indices of the PLN-dominated sample detected in the four Galactic BHXBs }
\label{minmax2}

\begin{tabular}{llcccc}
\toprule
          Source  &   Measurements  &              fractional rms (\%)  &         Obs ID &             index &         Obs ID \\
\midrule
\multirow{1}{*}{4U~1630$-$47} & minimum & ${1.50}_{-1.11}^{+2.29}$ & 60118-01-05-00 & ${-1.48}_{-0.16}^{+0.15}$ & 40418-01-18-05 \\ [5pt]
 & maximum & ${5.79}_{-0.57}^{+0.51}$ & 70417-01-07-02 & ${-0.69}_{-0.29}^{+0.22}$ & 90128-01-14-00 \\ [5pt]
\\
\multirow{1}{*}{H~1743$-$322} & minimum & ${1.73}_{-0.43}^{+0.49}$ & 90058-10-02-00 & ${-1.43}_{-0.18}^{+0.14}$ & 80146-01-75-01 \\ [5pt]
 & maximum & ${7.67}_{-0.49}^{+0.49}$ & 80146-01-22-01 & ${-0.79}_{-0.07}^{+0.07}$ & 80146-01-95-01 \\ [5pt]
\\
\multirow{1}{*}{XTE~J1817$-$330} & minimum & ${0.83}_{-0.46}^{+0.71}$ & 91110-02-06-01 & ${-1.64}_{-0.27}^{+0.25}$ & 91110-02-06-01 \\ [5pt]
 & maximum & ${2.70}_{-0.15}^{+0.15}$ & 91110-02-21-00 & ${-0.62}_{-0.15}^{+0.14}$ & 91110-02-22-00 \\ [5pt]
\\
\multirow{1}{*}{XTE~J1859$+$226} & minimum & ${0.89}_{-0.62}^{+0.66}$ & 40124-01-53-01 & ${-1.05}_{-1.41}^{+0.38}$ & 40124-01-53-01 \\ [5pt]
 & maximum & ${1.28}_{-0.60}^{+0.76}$ & 40124-01-53-02 & ${-0.77}_{-0.65}^{+0.32}$ & 40124-01-53-02 \\
\bottomrule
\\
\multicolumn{6}{l}{$Note.$ The uncertainties of the fractional rms and the PL indices correspond to the confidence level of 90\%}\\
\end{tabular}
\end{table*}

\begin{figure}
\centering
 \includegraphics[width=0.47\textwidth]{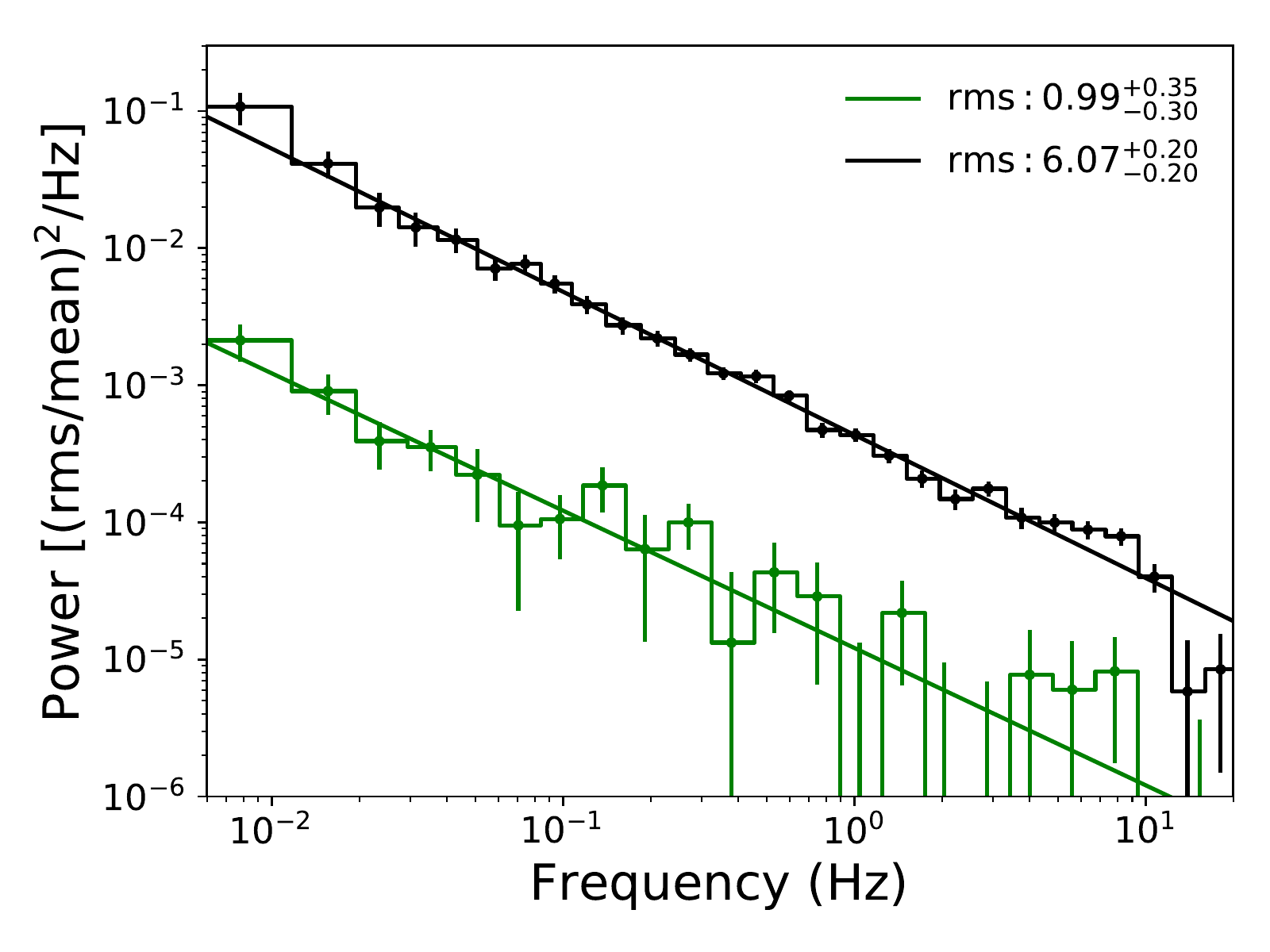}
\caption{Two representative RXTE/PCA power spectra in the soft state showing distinct fractional rms variability in the 0.01--20 Hz frequency range.}
\label{rms}
\end{figure}

As shown in Table \ref{minmax}, the maximum fractional rms is 5.53\% for 4U~1630$-$47, 6.07\% for H~1743$-$322, 2.70\% for XTE~J1817$-$330, and 1.28\% for XTE~J1859$+$226, respectively. The maximum fractional rms among the four sources roughly span by a factor of 4-5, which is surprisingly large. This is probably also larger than the span of the maximum rms in the black hole hard state among different sources. The minimum fractional rms of the soft states in the four sources are consistent with a value around 1\%, or even smaller, and the minimum was seen in XTE~J1817$-$330 which reached ${0.99}_{-0.30}^{+0.35}$\%. We plot the representative power spectra corresponding to the maximum and the minimum fractional rms in Figure \ref{rms}. The upper and lower limits on fractional rms correspond to a 90\% confidence interval. This yields an upper limit on the minimum fractional rms corresponding to the black hole soft state from the data we analyzed. 

Our measurements of the fractional rms variability in black hole soft states indicate that there is a systematic difference in the maximum fractional rms, the averaged fractional rms and the fractional rms of the median in the samples, and therefore differences are also in the fractional rms range, among the four black hole transients. The difference can only originate from distinct source properties. It is known that H~1743$-$322 and 4U~1630$-$47 are high inclination systems, XTE~J1859$+$226 has an intermediate inclination, while XTE~J1817$-$330 is a low inclination system \citep{2017MNRAS.464.2643V}. It is clear that the fractional rms variability of the high inclination systems seems significantly higher than that of the low inclination systems. Figure \ref{bb_rms} shows the averaged fractional rms of the PLN component integrated in 0.01-20 Hz of both the PLN-only sample and the PLN-dominated sample (see below) in relation to the averaged disk fraction, which is estimated by taking the average of the sample for each individual source, and the fractional rms of the median of the samples in relation to the disk fraction.

\subsubsection{Results of the PLN-dominated sample}
\begin{figure}
\centering
\includegraphics[width=0.45\textwidth]{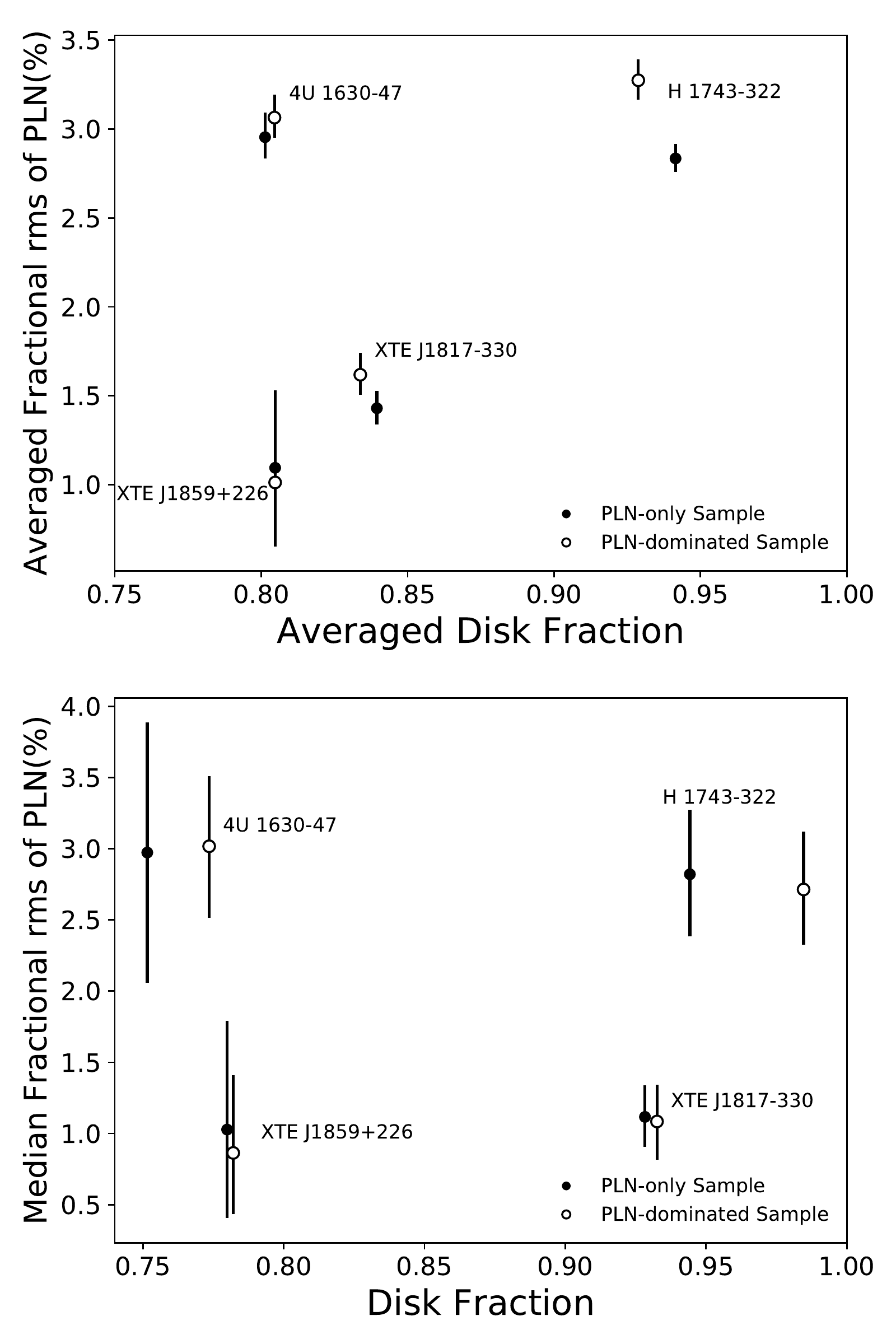}
\caption{Fractional rms variability vs. disk fraction relation. Upper panel: The averaged fractional rms of the PLN (0.01--20 Hz) against the averaged disk fraction of the PLN-only sample (filled circle) and PLN-dominated sample (open circle) of the X-ray observations of the soft spectral state for the four black hole transients. Both 4U~1630$-$47 and H~1743$-$322 belong to the high inclination sample, showing higher averaged fractional rms variability of the PLN. The maximum fractional rms of the PLN observed in these two sources are also the largest (see Table \ref{minmax}). Lower panel: The fractional rms of the median against the disk fraction of the PLN-only sample (filled circle) and PLN-dominated sample (open circle).}
\label{bb_rms}
\end{figure}

For the PLN-only sample, the BLN component might be present in the PDS. To estimate the contribution of potential BLN component and investigate the PLN component properties, we obtained an additional sample that is quoted as the PLN-dominated sample, of which the power spectra are in general composed of a dominant power-law plus a zero-centered Lorentzian. The maxima and minima of the PL indices and the fractional rms of the PLN of the PLN-dominated sample are listed in Table \ref{minmax2}. For the PLN-dominated sample, the maximum and the minimum of the PL indices were ${-0.62}_{-0.15}^{+0.14}$ and ${-1.64}_{-0.27}^{+0.25}$, respectively, with the errors correspond to 90\% confidence. We also calculated the PLN fractional rms by integrating the power-law in the frequency range between 0.01 and 20 Hz of the PLN-dominated sample. The fractional rms of the PLN component is on average around 2.9\% with the maximum at $7.67_{-0.49}^{+0.49}\%$ and minimum at $0.83_{-0.46}^{+0.71}\%$ in the sample. The maximum fractional rms of the PLN is 7.67\% for H~1743$-$322, 5.79\% for 4U~1630$-$47, 2.70\% for XTE~J1817$-$330, and 1.28\% for XTE~J1859$+$226, respectively. These also suggest that there is a distinction in the fractional rms of the PLN component among the four black hole transients. The relation between the PLN rms and the inclination angle still holds in the PLN-dominated sample (see Figure \ref{bb_rms}).

\subsubsection{The inclination dependence of the fractional variability}
\begin{figure}
\centering
 \includegraphics[width=0.45\textwidth]{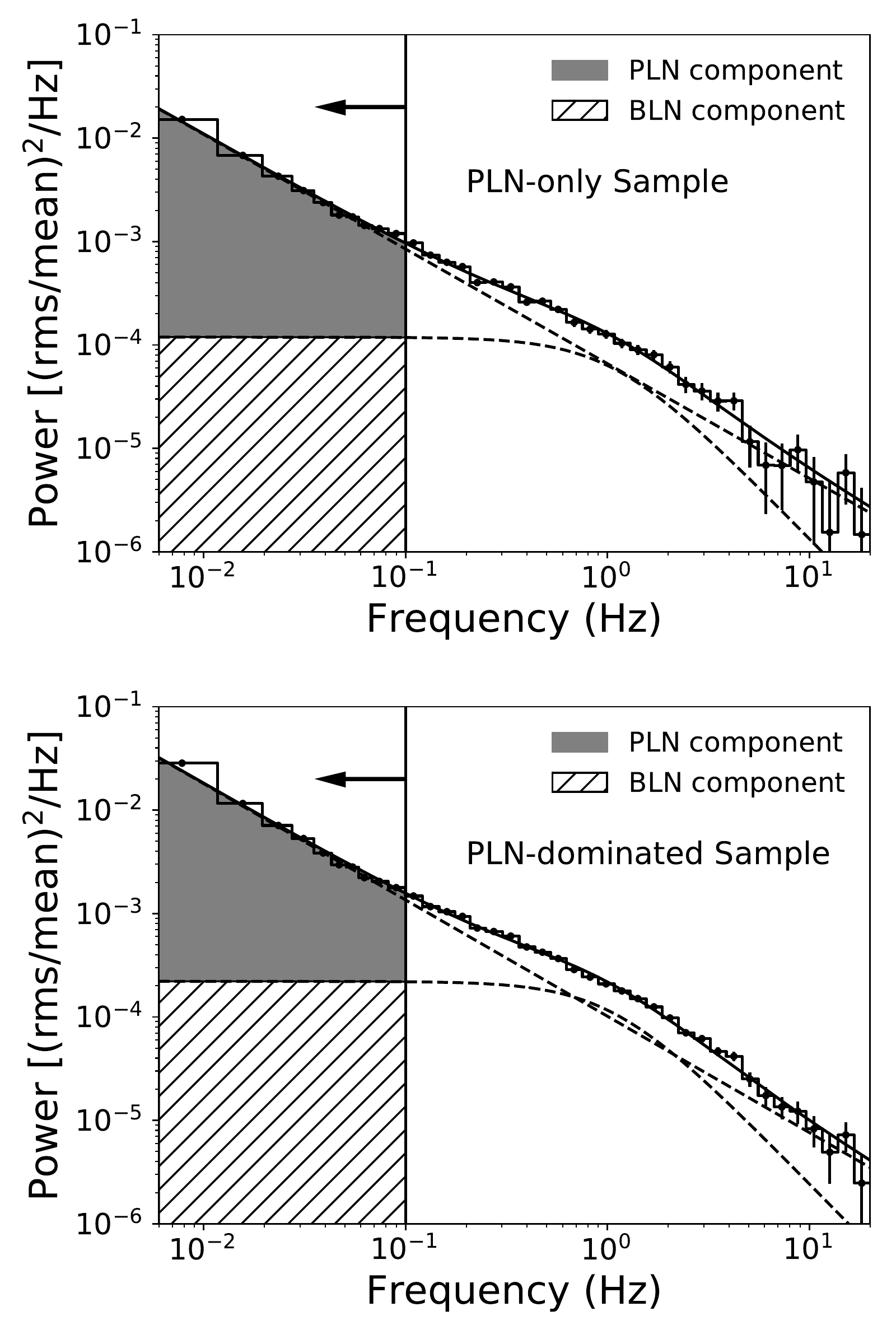}
\caption{The average power spectra of the PLN-only sample (upper panel) and the PLN-dominated sample (lower panel). The power spectra were fitted with the model composed of a single power-law for the PLN and a zero-centered Lorentzian for the BLN. The solid line represents the best-fit model and the two dash lines represent the two components in the model. The vertical solid line marks the frequency of 0.1 Hz. The power spectra below the frequency of 0.1 Hz are dominated by the PLN component (gray-shaded region) instead of the BLN component (straight-line-shaded region).}
\label{pds_avg}
\end{figure}

\begin{figure}
\centering
\includegraphics[width=0.45\textwidth]{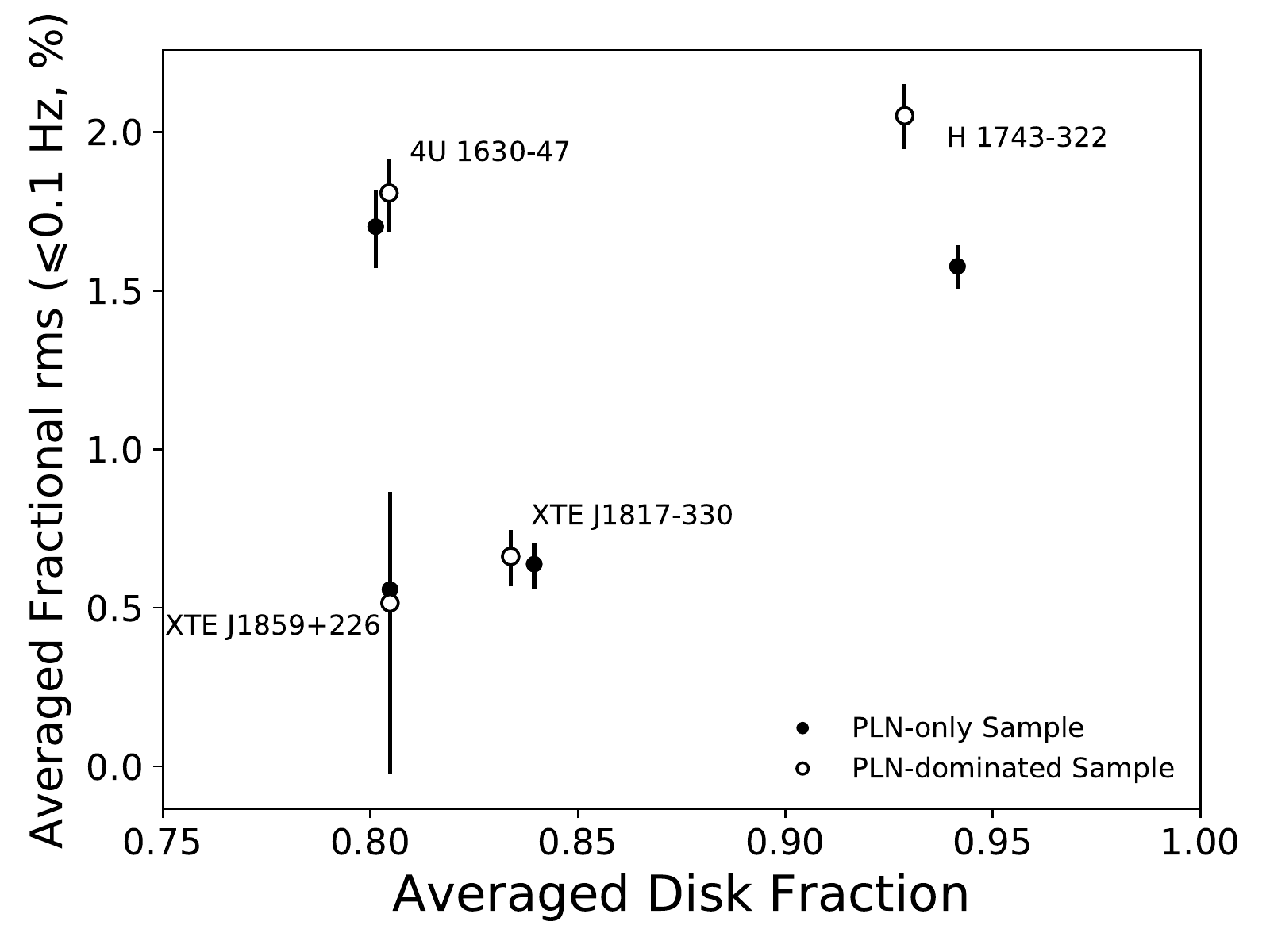}
\caption{The averaged fractional rms ($\leqslant$ 0.1 Hz) of the PLN against the averaged disk fraction of PLN-only sample (filled circle) and the PLN-dominated sample (open circle) of the X-ray observations corresponding to the soft spectral state for the four sources. The fractional rms calculated in the frequency below 0.1 Hz samples primarily the PLN. Similar to the results shown in Figure \ref{bb_rms}, the averaged fractional rms variability of the PLN of the high inclination systems is significantly larger.}
\label{bb_rms_01}
\end{figure}

To further estimate the rms variability of the PLN and eliminate the contribution of the BLN component, we have also obtained the average power spectra of the PLN-only sample and the PLN-dominated sample, respectively. We modeled the power spectra with a zero-centered Lorentzian for the BLN and a single power-law for the PLN. Figure \ref{pds_avg} shows the average power spectra of both samples. As we can see clearly, the power below 0.1 Hz does not contain a significant contribution from the BLN component. This is shown in Figure \ref{pds_avg}. The power contribution from the BLN component is represented by the straight-line-shaded region, while the power contribution from the PLN component is represented by the gray-shaded region. The area of the gray-shaded region is much larger than that of the straight-line-shaded region. This demonstrates that the power below 0.1 Hz describes the PLN component very well. Figure \ref{bb_rms_01} shows the averaged fractional rms by integrating the power-law below 0.1 Hz for the PLN-only sample and the PLN-dominated sample. As it shows, the fractional rms variability of the high inclination systems is significantly larger and the inclination-dependence of the fractional rms variability of the PLN is obvious. We noticed that a small number of samples were associated with small reduced $\chi^{2}$. We determined that the results are consistent with the same if excluding these samples. Our results tend to suggest an apparent positive correlation of the maximum fractional rms, the averaged fractional rms and the fractional rms of the median in the samples of the X-ray variability in the soft state with the inclination angle among the four black hole transients, confirming the inclination dependence of the fractional rms variability.

\section{Discussion and Conclusion}
The X-ray timing observations performed in a period of about 16 years by the {\it RXTE} have provided the most complete spectral and timing data archive to address statistical timing properties of those Galactic black hole transients. Here we present a statistical study of the power spectra corresponding to the soft X-ray spectral states in four of the Galactic black hole transients, namely 4U~J1630$-$47, H~1743$-$322, XTE~J1817$-$330, and XTE~J1859$+$226, in which at least twenty of PCA observations were performed to cover the soft energy spectral states. We found that the power spectra corresponding to the black hole soft state can be modeled with a PLN with the indices in a wide range between -1.64 and -0.62, while the fractional rms variability of the PLN is found in the range from 0.83\% to 7.67\%. 

In the persistent black hole binary Cygnus X$-$1, the index of its PLN was found around -1.0 in the Fourier frequency range of 0.125--256 Hz \citep{2014A&A...565A...1G}. The large range of the PL indices observed in the four black hole transients might indicate more variations in the black hole transients than those in the persistent Cygnus X$-$1. As seen in Figure \ref{index} and Figure \ref{rms}, the PL indices observed in the four black hole transients varied in a much larger range. This provides an important clue to the origin of the PL indices in black hole soft spectral states. 

\cite{1997MNRAS.292..679L} interpret the power spectrum of the emission from the standard disk as the inward propagation of modulation in $\dot{M}$, in which the standard disk acts as a filter which passes the $\dot{M}$ variations on time scales longer than the local viscous time $t_{\rm visc}(R)$ at any given radius $R$. Among the dynamical time scale, the thermal time scale and the viscous time scale in the accretion flow, the viscous time is the longest at a given radius, so that the frequency of the variation follows $f \sim t_{\rm visc}^{-1}(r)$. They assumed that the amplitude of the fluctuation of the viscosity parameter $\alpha$ depends on the radius $R$ as well, as $\sqrt{<\beta>} \propto r^{b}$, and then yields the corresponding power spectrum in the form $P(f) \propto f^{-1-4/3b}$. If the fluctuations of the viscosity $\alpha$ at all radii have the same relative amplitude, the power spectrum will display a PL index of -1. The variable PL indices of the power spectra observed in the four black hole transients then indicate that the fluctuation in the viscosity of the disk flow varied from observation to observation significantly even in the soft state of a single source. Notice that the PL noise variability with an index of -1 was successfully reproduced from this magnetohydrodynamic simulation of a geometrically thin accretion disk \citep{2016ApJ...826...40H}. In the same simulation, the fluctuation in the viscosity parameter $\alpha$ originates from the properties of the magnetic field in the accretion flow. Thus the PL index probably reflects the relation between the magnetic field and the radius $R$ in the accretion flow, making the magnetic field properties of the standard disk visible to our observers. 

Both 4U~1630$-$47 and H~1743$-$322 belong to the high inclination sample \citep{2017MNRAS.464.2643V} with the inclination angle up to 75 degrees \citep{2012ApJ...745L...7S}. Their maximum fractional rms variability of the PLN component is indeed at the high end, reached 7.67\% and 5.79\%, respectively (see Table \ref{minmax} and \ref{minmax2}). XTE~J1859$+$226 is probably of intermediate inclination \citep{2011MNRAS.413L..15C} and XTE~J1817$-$330 belongs to the low inclination sample based on direct measurements and indirect estimations \citep{2017MNRAS.464.2643V}. These two sources have much lower maximum fractional rms variability at around 2\% (see Table \ref{minmax} and \ref{minmax2} and Figure \ref{bb_rms} and \ref{bb_rms_01}). Our results then suggest that the fractional rms variability in black hole soft states is influenced by inclination effects. For the four sources in our sample, there is no significant correlation between the disk fraction and the PLN fractional rms, suggesting that the variation in the rms of the PLN is not caused by the variation of the disk flux fraction. As shown in Figure \ref{bb_rms} and \ref{bb_rms_01}, the averaged fractional rms as well as the fractional rms of the median of the soft state of the two high-inclination sources is larger, while their averaged disk fraction does not show systematic differences from that of the two sources with lower inclinations. The inclination dependence of the fractional rms variability of the black hole soft spectral states therefore is intrinsic to the PLN originated from the black hole accretion flow in the soft spectral state. 

Interestingly, as suggested by an early global magnetohydrodynamic simulation of an isothermal accretion disk around Schwarzschild black holes \citep{2003MNRAS.341.1041A}, a high inclination of the accretion disk would boost the power of variability at frequencies close to the inner disk edge. We have found signatures of an inclination dependence of the maximum fractional rms variability in our limited BHXB sample, but the inclination dependence is not entirely consistent with the theoretical simulation, as the boost of power in their simulation was found only corresponding to frequencies close to that of the inner disk edge. \cite{2013ApJ...770..135Y} studied the evolution of the black hole transient MAXI J1659$-$152 during its outburst and found that the PLN was associated with the thermal disk component no matter the source is in the soft state or the hard or intermediate state, suggesting that the PLN is a signature of the standard disk accretion flow. If this holds for all the other black hole transients, then the fractional rms variability of the PLN in the black hole soft state, intermediate state or even hard state in the soft X-ray band if detectable would signature the inclination of the optically thick disk component. The large differences in the maximum fractional rms variability (of a factor of four) among the four black hole transients indicate that the fractional rms is a sensitive probe of disk inclination in black hole soft state. Future statistical studies of the fractional rms variability of the PLN component in the disk spectral component during the soft state, intermediate state or even the hard state, would confirm the link between the maximum fractional rms variability (and thus the averaged fractional rms variability and the range of the fractional rms variability) and the inclination of the disk accretion flow.

\normalem
\begin{acknowledgements}
WY would like to acknowledge stimulating discussions with Professors Olmer Blaes, Marek Abramowicz, Jean-Pierre Lasota, Charles Gammie, Chris Reynolds, and Steven Balbus in the past. This work was supported in part by the National Program on Key Research and Development Project (Grant No. 2016YFA0400804) and the National Natural Science Foundation of China (Grant Nos. 11333005 and U1838203).  WY would also like to acknowledge the support by the FAST Scholar fellowship, which is supported by special funding for advanced users, budgeted and administrated by Center for Astronomical Mega-Science, Chinese Academy of Sciences (CAMS). We would like to acknowledge the use of {\it RXTE} standard data products and recipes from the {\it RXTE} Guest Observer Facility for data analysis, which is funded by NASA. This research also made use of data and/or software provided by the High Energy Astrophysics Science Archive Research Center (HEASARC). We thank the anonymous reviewer for constructive comments.
\end{acknowledgements}

\bibliographystyle{raa}
\bibliography{ms2021-0017.R1}

\begin{thebibliography}{54}
\providecommand\natexlab[1]{#1}
\providecommand\JournalTitle[1]{#1}

\bibitem[{Armitage} \& {Reynolds}(2003)]{2003MNRAS.341.1041A}
{Armitage}, P.~J., \& {Reynolds}, C.~S. 2003, \mnras, 341, 1041

\bibitem[{Balbus} \& {Hawley}(1991)]{1991ApJ...376..214B}
{Balbus}, S.~A., \& {Hawley}, J.~F. 1991, \apj, 376, 214

\bibitem[{Belloni} \& {Hasinger}(1990)]{1990A&A...227L..33B}
{Belloni}, T., \& {Hasinger}, G. 1990, \aap, 227, L33

\bibitem[{Belloni} {et~al.}(2005)]{2005A&A...440..207B}
{Belloni}, T., {Homan}, J., {Casella}, P., {et~al.} 2005, \aap, 440, 207

\bibitem[{Belloni}(2010)]{2010LNP...794...53B}
{Belloni}, T.~M. 2010, in Lecture Notes in Physics, Berlin Springer Verlag,
  Vol. 794, Lecture Notes in Physics, Berlin Springer Verlag, ed. T.~{Belloni},
  53

\bibitem[{Belloni} \& {Motta}(2016)]{2016ASSL..440...61B}
{Belloni}, T.~M., \& {Motta}, S.~E. 2016, in Astrophysics and Space Science
  Library, Vol. 440, Astrophysics of Black Holes: From Fundamental Aspects to
  Latest Developments, ed. C.~{Bambi}, 61

\bibitem[{Belloni} {et~al.}(2011)]{2011BASI...39..409B}
{Belloni}, T.~M., {Motta}, S.~E., \& {Mu{\~n}oz-Darias}, T. 2011, Bulletin of
  the Astronomical Society of India, 39, 409

\bibitem[{Belloni} {et~al.}(2002)]{2002ApJ...572..392B}
{Belloni}, T., {Psaltis}, D., \& {van der Klis}, M. 2002, \apj, 572, 392

\bibitem[{Bowyer} {et~al.}(1965)]{1965Sci...147..394B}
{Bowyer}, S., {Byram}, E.~T., {Chubb}, T.~A., \& {Friedman}, H. 1965, Science,
  147, 394

\bibitem[{Capitanio} {et~al.}(2009)]{2009MNRAS.398.1194C}
{Capitanio}, F., {Belloni}, T., {Del Santo}, M., \& {Ubertini}, P. 2009,
  \mnras, 398, 1194

\bibitem[{Chen} {et~al.}(1997)]{1997ApJ...491..312C}
{Chen}, W., {Shrader}, C.~R., \& {Livio}, M. 1997, \apj, 491, 312

\bibitem[{Corral-Santana} {et~al.}(2011)]{2011MNRAS.413L..15C}
{Corral-Santana}, J.~M., {Casares}, J., {Shahbaz}, T., {et~al.} 2011, \mnras,
  413, L15

\bibitem[{Cui} {et~al.}(1997)]{1997ApJ...474L..57C}
{Cui}, W., {Heindl}, W.~A., {Rothschild}, R.~E., {et~al.} 1997, \apjl, 474, L57

\bibitem[{Dove} {et~al.}(1998)]{1998MNRAS.298..729D}
{Dove}, J.~B., {Wilms}, J., {Nowak}, M.~A., {Vaughan}, B.~A., \& {Begelman},
  M.~C. 1998, \mnras, 298, 729

\bibitem[{Elvis} {et~al.}(1975)]{1975Natur.257..656E}
{Elvis}, M., {Page}, C.~G., {Pounds}, K.~A., {Ricketts}, M.~J., \& {Turner},
  M.~J.~L. 1975, \nat, 257, 656

\bibitem[{Esin} {et~al.}(1997)]{1997ApJ...489..865E}
{Esin}, A.~A., {McClintock}, J.~E., \& {Narayan}, R. 1997, \apj, 489, 865

\bibitem[{Frank} {et~al.}(2002)]{2002apa..book.....F}
{Frank}, J., {King}, A., \& {Raine}, D.~J. 2002, {Accretion Power in
  Astrophysics: Third Edition}

\bibitem[{Gierli{\'n}ski} {et~al.}(1999)]{1999MNRAS.309..496G}
{Gierli{\'n}ski}, M., {Zdziarski}, A.~A., {Poutanen}, J., {et~al.} 1999,
  \mnras, 309, 496

\bibitem[{Gilfanov}(2010)]{2010LNP...794...17G}
{Gilfanov}, M. 2010, in Lecture Notes in Physics, Berlin Springer Verlag, Vol.
  794, Lecture Notes in Physics, Berlin Springer Verlag, ed. T.~{Belloni}, 17

\bibitem[{Gilfanov} {et~al.}(2000)]{2000MNRAS.316..923G}
{Gilfanov}, M., {Churazov}, E., \& {Revnivtsev}, M. 2000, \mnras, 316, 923

\bibitem[{Grinberg} {et~al.}(2014)]{2014A&A...565A...1G}
{Grinberg}, V., {Pottschmidt}, K., {B{\"o}ck}, M., {et~al.} 2014, \aap, 565, A1

\bibitem[{Hogg} \& {Reynolds}(2016)]{2016ApJ...826...40H}
{Hogg}, J.~D., \& {Reynolds}, C.~S. 2016, \apj, 826, 40

\bibitem[{Homan} {et~al.}(2001)]{2001ApJS..132..377H}
{Homan}, J., {Wijnands}, R., {van der Klis}, M., {et~al.} 2001, \apjs, 132, 377

\bibitem[{Hynes} {et~al.}(2002)]{2002MNRAS.331..169H}
{Hynes}, R.~I., {Haswell}, C.~A., {Chaty}, S., {Shrader}, C.~R., \& {Cui}, W.
  2002, \mnras, 331, 169

\bibitem[{Jahoda} {et~al.}(2006)]{2006ApJS..163..401J}
{Jahoda}, K., {Markwardt}, C.~B., {Radeva}, Y., {et~al.} 2006, \apjs, 163, 401

\bibitem[{Kaluzienski} {et~al.}(1977)]{1977IAUC.3106....3K}
{Kaluzienski}, L.~J., {Holt}, S.~S., \& {Glass}, I.~S. 1977, \iaucirc, 3106

\bibitem[{Klein-Wolt}(2004)]{2004PhDT.......415K}
{Klein-Wolt}, M. 2004, {Black Hole X-ray Binaries}, PhD thesis, University of
  Amsterdam

\bibitem[{Klein-Wolt} \& {van der Klis}(2008)]{2008ApJ...675.1407K}
{Klein-Wolt}, M., \& {van der Klis}, M. 2008, \apj, 675, 1407

\bibitem[{Kuulkers} {et~al.}(1997)]{1997MNRAS.291...81K}
{Kuulkers}, E., {Parmar}, A.~N., {Kitamoto}, S., {Cominsky}, L.~R., \& {Sood},
  R.~K. 1997, \mnras, 291, 81

\bibitem[{Lamers} {et~al.}(1976)]{1976A&A....49..327L}
{Lamers}, H.~J.~G.~L.~M., {van den Heuvel}, E.~P.~J., \& {Petterson}, J.~A.
  1976, \aap, 49, 327

\bibitem[{Leahy} {et~al.}(1983)]{1983ApJ...266..160L}
{Leahy}, D.~A., {Darbro}, W., {Elsner}, R.~F., {et~al.} 1983, \apj, 266, 160

\bibitem[{Lyubarskii}(1997)]{1997MNRAS.292..679L}
{Lyubarskii}, Y.~E. 1997, \mnras, 292, 679

\bibitem[{McClintock} {et~al.}(2009)]{2009ApJ...698.1398M}
{McClintock}, J.~E., {Remillard}, R.~A., {Rupen}, M.~P., {et~al.} 2009, \apj,
  698, 1398

\bibitem[{Miyamoto} {et~al.}(1993)]{1993ApJ...403L..39M}
{Miyamoto}, S., {Iga}, S., {Kitamoto}, S., \& {Kamado}, Y. 1993, \apjl, 403,
  L39

\bibitem[{Miyamoto} {et~al.}(1994)]{1994ApJ...435..398M}
{Miyamoto}, S., {Kitamoto}, S., {Iga}, S., {Hayashida}, K., \& {Terada}, K.
  1994, \apj, 435, 398

\bibitem[{Miyamoto} {et~al.}(1992)]{1992ApJ...391L..21M}
{Miyamoto}, S., {Kitamoto}, S., {Iga}, S., {Negoro}, H., \& {Terada}, K. 1992,
  \apjl, 391, L21

\bibitem[{Motta} {et~al.}(2010)]{2010MNRAS.408.1796M}
{Motta}, S., {Mu{\~n}oz-Darias}, T., \& {Belloni}, T. 2010, \mnras, 408, 1796

\bibitem[{Motta} {et~al.}(2011)]{2011MNRAS.418.2292M}
{Motta}, S., {Mu{\~n}oz-Darias}, T., {Casella}, P., {Belloni}, T., \& {Homan},
  J. 2011, \mnras, 418, 2292

\bibitem[{Nowak} {et~al.}(1999)]{1999ApJ...510..874N}
{Nowak}, M.~A., {Vaughan}, B.~A., {Wilms}, J., {Dove}, J.~B., \& {Begelman},
  M.~C. 1999, \apj, 510, 874

\bibitem[{Polyakov} {et~al.}(2012)]{2012AJ....143..148P}
{Polyakov}, Y.~S., {Neilsen}, J., \& {Timashev}, S.~F. 2012, \aj, 143, 148

\bibitem[{Remillard} \& {McClintock}(2006)]{2006ARA&A..44...49R}
{Remillard}, R.~A., \& {McClintock}, J.~E. 2006, \araa, 44, 49

\bibitem[{Remillard} {et~al.}(2006)]{2006ATel..714....1R}
{Remillard}, R., {Levine}, A.~M., {Morgan}, E.~H., {Markwardt}, C.~B., \&
  {Swank}, J.~H. 2006, The Astronomer's Telegram, 714

\bibitem[{Shakura} \& {Sunyaev}(1973)]{1973A&A....24..337S}
{Shakura}, N.~I., \& {Sunyaev}, R.~A. 1973, \aap, 24, 337

\bibitem[{Steiner} {et~al.}(2012)]{2012ApJ...745L...7S}
{Steiner}, J.~F., {McClintock}, J.~E., \& {Reid}, M.~J. 2012, \apjl, 745, L7

\bibitem[{Stiele} \& {Yu}(2015)]{2015MNRAS.452.3666S}
{Stiele}, H., \& {Yu}, W. 2015, \mnras, 452, 3666

\bibitem[{Tananbaum} {et~al.}(1972)]{1972ApJ...177L...5T}
{Tananbaum}, H., {Gursky}, H., {Kellogg}, E., {Giacconi}, R., \& {Jones}, C.
  1972, \apjl, 177, L5

\bibitem[{van den Eijnden} {et~al.}(2017)]{2017MNRAS.464.2643V}
{van den Eijnden}, J., {Ingram}, A., {Uttley}, P., {et~al.} 2017, \mnras, 464,
  2643

\bibitem[{van der Klis}(1989)]{1989ARA&A..27..517V}
{van der Klis}, M. 1989, \araa, 27, 517

\bibitem[{Wijnands} \& {van der Klis}(1999)]{1999ApJ...514..939W}
{Wijnands}, R., \& {van der Klis}, M. 1999, \apj, 514, 939

\bibitem[{Yan} \& {Yu}(2015)]{2015ApJ...805...87Y}
{Yan}, Z., \& {Yu}, W. 2015, \apj, 805, 87

\bibitem[{Yu} \& {Yan}(2009)]{2009ApJ...701.1940Y}
{Yu}, W., \& {Yan}, Z. 2009, \apj, 701, 1940

\bibitem[{Yu} \& {Zhang}(2013)]{2013ApJ...770..135Y}
{Yu}, W., \& {Zhang}, W. 2013, \apj, 770, 135

\bibitem[{Zhang} {et~al.}(1997)]{1997ApJ...477L..95Z}
{Zhang}, S.~N., {Cui}, W., {Harmon}, B.~A., {et~al.} 1997, \apjl, 477, L95

\bibitem[{Zhang} {et~al.}(1995)]{1995ApJ...449..930Z}
{Zhang}, W., {Jahoda}, K., {Swank}, J.~H., {Morgan}, E.~H., \& {Giles}, A.~B.
  1995, \apj, 449, 930

\end{thebibliography}

\onecolumn
\begin{longtable}{lcccccccc}
\caption{The spectral fits to the X-ray power spectra for four black hole transients} \label{fitting result}\\
\toprule
ObsID & $\rm {F_{disk}}^a$ & $\rm index^{b}$ & $\rm {N (\times10^{-4}})^{c}$ & rms (\%) & $\chi_{\nu}^{2}$ & $dof$ \\ 
\midrule 
\endfirsthead

\toprule
ObsID & $\rm {F_{disk}}^a$ & $\rm index^{b}$ & $\rm {N (\times10^{-4}})^{c}$ & rms (\%) & $\chi_{\nu}^{2}$ & $dof$ \\ 
\midrule 
\endhead

\bottomrule
\endfoot

\\
\endlastfoot

\multicolumn{7}{c}{4U~1630$-$47} \\
\midrule
10411-01-08-00 & 0.88 & ${-0.86}_{-0.03}^{+0.03}$ & ${2.53}_{-0.18}^{+0.18}$ & ${4.31}_{-0.15}^{+0.15}$ & 1.60 & 20 \\ [5pt]
10411-01-09-00 & 0.87 & ${-0.82}_{-0.04}^{+0.04}$ & ${1.93}_{-0.18}^{+0.18}$ & ${3.75}_{-0.17}^{+0.17}$ & 1.92 & 20 \\ [5pt]
10411-01-13-00 & 0.87 & ${-0.88}_{-0.03}^{+0.03}$ & ${2.77}_{-0.20}^{+0.20}$ & ${4.53}_{-0.16}^{+0.16}$ & 1.73 & 20 \\ [5pt]
40418-01-17-02 & 0.76 & ${-1.24}_{-0.30}^{+0.22}$ & ${0.38}_{-0.25}^{+0.35}$ & ${2.15}_{-0.99}^{+1.12}$ & 0.97 & 20 \\ [5pt]
40418-01-18-00 & 0.76 & ${-1.40}_{-0.21}^{+0.18}$ & ${0.26}_{-0.14}^{+0.20}$ & ${2.14}_{-0.86}^{+1.01}$ & 1.10 & 20 \\ [5pt]
40418-01-18-02 & 0.80 & ${-1.20}_{-0.28}^{+0.21}$ & ${0.38}_{-0.24}^{+0.33}$ & ${2.05}_{-0.86}^{+0.99}$ & 0.44 & 20 \\ [5pt]
40418-01-18-05 & 0.77 & ${-1.48}_{-0.16}^{+0.15}$ & ${0.17}_{-0.08}^{+0.11}$ & ${1.96}_{-0.66}^{+0.75}$ & 0.62 & 20 \\ [5pt]
60118-01-05-00 & 0.83 & ${-1.16}_{-1.43}^{+0.47}$ & ${0.22}_{-0.22}^{+0.65}$ & ${1.50}_{-1.11}^{+2.29}$ & 1.29 & 9 \\ [5pt]
60118-01-07-00 & 0.76 & ${-1.22}_{-0.22}^{+0.18}$ & ${0.31}_{-0.16}^{+0.21}$ & ${1.90}_{-0.66}^{+0.72}$ & 0.88 & 20 \\ [5pt]
60118-01-09-00 & 0.79 & ${-1.12}_{-0.13}^{+0.12}$ & ${0.39}_{-0.16}^{+0.20}$ & ${1.92}_{-0.45}^{+0.52}$ & 1.39 & 20 \\ [5pt]
60118-01-10-00 & 0.89 & ${-1.09}_{-0.22}^{+0.18}$ & ${0.43}_{-0.25}^{+0.31}$ & ${1.97}_{-0.66}^{+0.76}$ & 0.89 & 20 \\ [5pt]
60118-01-11-00 & 0.88 & ${-1.05}_{-0.14}^{+0.12}$ & ${0.50}_{-0.19}^{+0.21}$ & ${2.07}_{-0.43}^{+0.46}$ & 0.92 & 20 \\ [5pt]
60118-01-14-00 & 0.86 & ${-1.14}_{-0.46}^{+0.27}$ & ${0.25}_{-0.19}^{+0.32}$ & ${1.56}_{-0.88}^{+1.08}$ & 1.21 & 20 \\ [5pt]
70113-01-32-00 & 0.82 & ${-1.16}_{-0.53}^{+0.36}$ & ${1.37}_{-1.18}^{+2.17}$ & ${3.73}_{-2.43}^{+3.21}$ & 0.57 & 9 \\ [5pt]
70113-01-35-00 & 0.83 & ${-1.31}_{-0.17}^{+0.17}$ & ${1.27}_{-0.56}^{+0.81}$ & ${4.22}_{-1.29}^{+1.61}$ & 1.12 & 20 \\ [5pt]
70113-01-40-00 & 0.77 & ${-0.86}_{-0.11}^{+0.10}$ & ${4.16}_{-1.32}^{+1.37}$ & ${5.53}_{-0.88}^{+0.91}$ & 1.07 & 20 \\ [5pt]
80420-01-16-01 & 0.75 & ${-1.07}_{-0.23}^{+0.17}$ & ${1.01}_{-0.53}^{+0.58}$ & ${2.97}_{-0.92}^{+0.91}$ & 0.74 & 9 \\ [5pt]
80420-01-17-00 & 0.79 & ${-1.00}_{-0.08}^{+0.07}$ & ${1.33}_{-0.25}^{+0.25}$ & ${3.26}_{-0.34}^{+0.33}$ & 1.18 & 20 \\ [5pt]
80420-01-17-01 & 0.76 & ${-1.40}_{-0.17}^{+0.16}$ & ${0.40}_{-0.17}^{+0.23}$ & ${2.69}_{-0.86}^{+0.98}$ & 0.70 & 20 \\ [5pt]
80420-01-18-00 & 0.76 & ${-1.16}_{-0.10}^{+0.09}$ & ${1.02}_{-0.23}^{+0.23}$ & ${3.22}_{-0.46}^{+0.46}$ & 0.88 & 20 \\ [5pt]
90128-01-14-00 & 0.87 & ${-0.69}_{-0.29}^{+0.22}$ & ${0.87}_{-0.52}^{+0.47}$ & ${2.55}_{-0.79}^{+0.71}$ & 0.51 & 20 \\ [5pt]
90128-01-16-00 & 0.86 & ${-1.02}_{-0.28}^{+0.21}$ & ${0.45}_{-0.29}^{+0.37}$ & ${1.93}_{-0.70}^{+0.83}$ & 1.01 & 20 \\ [5pt]
90410-01-01-00 & 0.91 & ${-0.98}_{-0.16}^{+0.13}$ & ${0.65}_{-0.29}^{+0.31}$ & ${2.26}_{-0.53}^{+0.56}$ & 0.95 & 20 \\ [5pt]
90410-01-01-01 & 0.84 & ${-1.00}_{-0.10}^{+0.09}$ & ${1.16}_{-0.34}^{+0.36}$ & ${3.05}_{-0.48}^{+0.50}$ & 0.70 & 20 \\ [5pt]
90410-01-03-00 & 0.77 & ${-1.07}_{-0.12}^{+0.11}$ & ${0.91}_{-0.29}^{+0.32}$ & ${2.84}_{-0.50}^{+0.53}$ & 0.60 & 20 \\ [5pt]
90410-01-03-01 & 0.78 & ${-0.95}_{-0.10}^{+0.09}$ & ${0.92}_{-0.27}^{+0.28}$ & ${2.67}_{-0.40}^{+0.42}$ & 1.26 & 20 \\ [5pt]
90410-01-04-01 & 0.79 & ${-0.76}_{-0.12}^{+0.10}$ & ${1.25}_{-0.42}^{+0.42}$ & ${3.02}_{-0.51}^{+0.51}$ & 0.82 & 20 \\ [5pt]
90410-01-04-02 & 0.75 & ${-0.77}_{-0.16}^{+0.14}$ & ${0.86}_{-0.37}^{+0.37}$ & ${2.51}_{-0.54}^{+0.54}$ & 1.34 & 20 \\ [5pt]
90410-01-04-03 & 0.78 & ${-0.97}_{-0.41}^{+0.23}$ & ${0.73}_{-0.57}^{+0.64}$ & ${2.39}_{-1.03}^{+1.08}$ & 1.51 & 20 \\ [5pt]
90410-01-04-04 & 0.77 & ${-0.81}_{-0.12}^{+0.11}$ & ${1.25}_{-0.41}^{+0.41}$ & ${3.02}_{-0.50}^{+0.49}$ & 0.87 & 20 \\ [5pt]
90410-01-05-01 & 0.79 & ${-0.84}_{-0.11}^{+0.09}$ & ${1.29}_{-0.36}^{+0.36}$ & ${3.07}_{-0.43}^{+0.43}$ & 1.16 & 20 \\ [5pt]
90410-01-06-00 & 0.76 & ${-0.93}_{-0.07}^{+0.07}$ & ${1.09}_{-0.20}^{+0.20}$ & ${2.87}_{-0.27}^{+0.27}$ & 1.23 & 20 \\ [5pt]
90410-01-06-01 & 0.76 & ${-1.00}_{-0.08}^{+0.07}$ & ${1.13}_{-0.24}^{+0.25}$ & ${3.00}_{-0.34}^{+0.35}$ & 1.58 & 20 \\ [5pt]
90410-01-07-00 & 0.78 & ${-0.95}_{-0.07}^{+0.07}$ & ${1.37}_{-0.26}^{+0.27}$ & ${3.26}_{-0.32}^{+0.33}$ & 1.31 & 20 \\ [5pt]
90410-01-07-01 & 0.75 & ${-0.90}_{-0.13}^{+0.11}$ & ${1.29}_{-0.44}^{+0.46}$ & ${3.11}_{-0.54}^{+0.56}$ & 0.60 & 20 \\ [5pt]
90410-01-08-00 & 0.79 & ${-0.92}_{-0.04}^{+0.04}$ & ${1.49}_{-0.15}^{+0.16}$ & ${3.35}_{-0.18}^{+0.18}$ & 1.84 & 20 \\ [5pt]
90410-01-09-00 & 0.81 & ${-0.97}_{-0.14}^{+0.12}$ & ${1.29}_{-0.40}^{+0.39}$ & ${3.17}_{-0.52}^{+0.51}$ & 1.02 & 20 \\ [5pt]
90410-01-10-02 & 0.80 & ${-0.87}_{-0.06}^{+0.06}$ & ${2.46}_{-0.35}^{+0.34}$ & ${4.26}_{-0.31}^{+0.30}$ & 0.59 & 20 \\ [5pt]
91704-03-06-00 & 0.78 & ${-0.84}_{-0.04}^{+0.04}$ & ${2.30}_{-0.20}^{+0.20}$ & ${4.10}_{-0.18}^{+0.18}$ & 1.58 & 20 \\ [5pt]
91704-03-14-01 & 0.77 & ${-0.81}_{-0.04}^{+0.04}$ & ${2.34}_{-0.23}^{+0.23}$ & ${4.13}_{-0.20}^{+0.20}$ & 1.33 & 20 \\ [5pt]
91704-03-17-02 & 0.75 & ${-1.02}_{-0.08}^{+0.08}$ & ${1.44}_{-0.29}^{+0.30}$ & ${3.44}_{-0.38}^{+0.38}$ & 0.81 & 20 \\ [5pt]
91704-03-31-02 & 0.76 & ${-0.81}_{-0.14}^{+0.12}$ & ${0.96}_{-0.38}^{+0.37}$ & ${2.64}_{-0.52}^{+0.52}$ & 1.51 & 9 \\ [5pt]
\midrule
\multicolumn{7}{c}{H~1743$-$322} \\
\midrule
80135-02-04-00 & 0.99 & ${-0.87}_{-0.06}^{+0.06}$ & ${1.04}_{-0.17}^{+0.17}$ & ${2.77}_{-0.23}^{+0.23}$ & 1.55 & 20 \\ [5pt]
80137-01-01-00 & 0.92 & ${-1.04}_{-0.17}^{+0.15}$ & ${0.37}_{-0.17}^{+0.21}$ & ${1.76}_{-0.44}^{+0.52}$ & 1.10 & 20 \\ [5pt]
80137-01-02-00 & 0.89 & ${-0.97}_{-0.12}^{+0.11}$ & ${0.56}_{-0.21}^{+0.23}$ & ${2.10}_{-0.40}^{+0.45}$ & 0.75 & 20 \\ [5pt]
80137-01-06-00 & 0.92 & ${-1.09}_{-0.20}^{+0.16}$ & ${0.37}_{-0.20}^{+0.26}$ & ${1.84}_{-0.57}^{+0.68}$ & 0.81 & 20 \\ [5pt]
80137-01-10-00 & 0.83 & ${-0.85}_{-0.48}^{+0.24}$ & ${0.42}_{-0.37}^{+0.48}$ & ${1.76}_{-0.77}^{+0.99}$ & 0.41 & 20 \\ [5pt]
80137-01-11-00 & 0.80 & ${-0.94}_{-0.13}^{+0.10}$ & ${0.59}_{-0.23}^{+0.24}$ & ${2.12}_{-0.42}^{+0.44}$ & 1.87 & 20 \\ [5pt]
80144-01-03-00 & 0.77 & ${-1.04}_{-0.03}^{+0.03}$ & ${4.35}_{-0.23}^{+0.22}$ & ${6.07}_{-0.20}^{+0.20}$ & 1.82 & 20 \\ [5pt]
80146-01-01-10 & 0.98 & ${-0.90}_{-0.07}^{+0.06}$ & ${1.10}_{-0.22}^{+0.22}$ & ${2.86}_{-0.29}^{+0.29}$ & 1.08 & 20 \\ [5pt]
80146-01-01-11 & 0.98 & ${-0.95}_{-0.09}^{+0.09}$ & ${0.96}_{-0.26}^{+0.28}$ & ${2.71}_{-0.39}^{+0.41}$ & 0.92 & 20 \\ [5pt]
80146-01-04-10 & 0.93 & ${-0.88}_{-0.25}^{+0.19}$ & ${0.74}_{-0.42}^{+0.43}$ & ${2.34}_{-0.66}^{+0.69}$ & 0.75 & 20 \\ [5pt]
80146-01-05-10 & 0.98 & ${-0.87}_{-0.12}^{+0.10}$ & ${1.15}_{-0.36}^{+0.36}$ & ${2.91}_{-0.46}^{+0.46}$ & 0.73 & 20 \\ [5pt]
80146-01-06-10 & 0.96 & ${-0.98}_{-0.12}^{+0.11}$ & ${1.05}_{-0.36}^{+0.39}$ & ${2.88}_{-0.52}^{+0.54}$ & 0.88 & 20 \\ [5pt]
80146-01-07-10 & 0.98 & ${-0.87}_{-0.11}^{+0.08}$ & ${1.32}_{-0.35}^{+0.30}$ & ${3.11}_{-0.42}^{+0.35}$ & 1.57 & 20 \\ [5pt]
80146-01-12-10 & 0.97 & ${-0.87}_{-0.08}^{+0.07}$ & ${1.49}_{-0.32}^{+0.32}$ & ${3.32}_{-0.36}^{+0.36}$ & 1.53 & 20 \\ [5pt]
80146-01-15-10 & 0.95 & ${-1.11}_{-0.29}^{+0.22}$ & ${0.53}_{-0.37}^{+0.53}$ & ${2.22}_{-0.94}^{+1.18}$ & 1.04 & 20 \\ [5pt]
80146-01-70-00 & 0.78 & ${-1.18}_{-0.03}^{+0.03}$ & ${2.06}_{-0.13}^{+0.13}$ & ${4.64}_{-0.20}^{+0.20}$ & 1.28 & 20 \\ [5pt]
80146-01-82-00 & 0.96 & ${-1.02}_{-0.05}^{+0.05}$ & ${1.76}_{-0.18}^{+0.18}$ & ${3.81}_{-0.22}^{+0.22}$ & 1.53 & 20 \\ [5pt]
80146-01-82-01 & 0.97 & ${-0.96}_{-0.07}^{+0.07}$ & ${1.49}_{-0.28}^{+0.29}$ & ${3.39}_{-0.33}^{+0.34}$ & 1.76 & 20 \\ [5pt]
80146-01-84-00 & 0.98 & ${-0.87}_{-0.05}^{+0.05}$ & ${1.53}_{-0.18}^{+0.18}$ & ${3.36}_{-0.20}^{+0.20}$ & 1.79 & 20 \\ [5pt]
80146-01-87-00 & 0.99 & ${-0.81}_{-0.06}^{+0.06}$ & ${1.22}_{-0.19}^{+0.19}$ & ${2.98}_{-0.24}^{+0.24}$ & 1.39 & 20 \\ [5pt]
80146-01-92-00 & 0.99 & ${-0.77}_{-0.06}^{+0.06}$ & ${1.29}_{-0.21}^{+0.21}$ & ${3.08}_{-0.25}^{+0.25}$ & 1.27 & 20 \\ [5pt]
80146-01-92-01 & 0.99 & ${-0.90}_{-0.05}^{+0.05}$ & ${1.07}_{-0.15}^{+0.15}$ & ${2.83}_{-0.20}^{+0.20}$ & 1.80 & 20 \\ [5pt]
80146-01-95-00 & 0.98 & ${-0.85}_{-0.05}^{+0.05}$ & ${0.97}_{-0.14}^{+0.15}$ & ${2.67}_{-0.20}^{+0.20}$ & 1.73 & 20 \\ [5pt]
80146-01-95-01 & 0.99 & ${-0.79}_{-0.07}^{+0.07}$ & ${1.13}_{-0.22}^{+0.22}$ & ${2.86}_{-0.28}^{+0.28}$ & 1.02 & 20 \\ [5pt]
80146-01-97-00 & 0.98 & ${-1.01}_{-0.05}^{+0.05}$ & ${1.45}_{-0.17}^{+0.18}$ & ${3.43}_{-0.23}^{+0.23}$ & 1.04 & 20 \\ [5pt]
80146-01-98-00 & 0.96 & ${-1.07}_{-0.08}^{+0.07}$ & ${1.36}_{-0.18}^{+0.18}$ & ${3.46}_{-0.30}^{+0.30}$ & 1.51 & 20 \\ [5pt]
90058-10-02-00 & 0.94 & ${-0.94}_{-0.17}^{+0.14}$ & ${0.39}_{-0.19}^{+0.22}$ & ${1.73}_{-0.43}^{+0.49}$ & 0.59 & 20 \\ [5pt]
90058-10-04-00 & 0.95 & ${-0.92}_{-0.10}^{+0.09}$ & ${0.67}_{-0.20}^{+0.22}$ & ${2.25}_{-0.35}^{+0.37}$ & 1.55 & 20 \\ [5pt]
90058-10-08-01 & 0.92 & ${-1.07}_{-0.18}^{+0.16}$ & ${0.43}_{-0.21}^{+0.26}$ & ${1.94}_{-0.53}^{+0.62}$ & 1.18 & 20 \\ [5pt]
90421-01-01-00 & 0.94 & ${-0.90}_{-0.11}^{+0.10}$ & ${1.07}_{-0.33}^{+0.34}$ & ${2.82}_{-0.44}^{+0.45}$ & 1.20 & 20 \\ [5pt]
90421-01-01-01 & 0.96 & ${-0.95}_{-0.09}^{+0.08}$ & ${0.82}_{-0.22}^{+0.22}$ & ${2.50}_{-0.35}^{+0.34}$ & 1.70 & 20 \\ [5pt]
90421-01-02-00 & 0.96 & ${-0.94}_{-0.09}^{+0.08}$ & ${0.91}_{-0.23}^{+0.23}$ & ${2.63}_{-0.34}^{+0.34}$ & 1.28 & 20 \\ [5pt]
90421-01-03-00 & 0.95 & ${-1.24}_{-0.20}^{+0.17}$ & ${0.44}_{-0.21}^{+0.26}$ & ${2.29}_{-0.75}^{+0.82}$ & 0.45 & 20 \\ [5pt]
90421-01-04-00 & 0.96 & ${-0.96}_{-0.14}^{+0.13}$ & ${1.11}_{-0.37}^{+0.39}$ & ${2.93}_{-0.52}^{+0.54}$ & 0.54 & 20 \\ [5pt]
\midrule
\multicolumn{7}{c}{XTE~J1817$-$330} \\
\midrule
91110-02-04-00 & 0.79 & ${-1.03}_{-0.14}^{+0.13}$ & ${0.12}_{-0.04}^{+0.05}$ & ${1.00}_{-0.20}^{+0.22}$ & 1.06 & 9 \\ [5pt]
91110-02-06-01 & 0.77 & ${-1.03}_{-0.10}^{+0.09}$ & ${0.16}_{-0.04}^{+0.04}$ & ${1.16}_{-0.15}^{+0.15}$ & 1.95 & 20 \\ [5pt]
91110-02-11-00 & 0.93 & ${-0.78}_{-0.14}^{+0.12}$ & ${0.17}_{-0.06}^{+0.07}$ & ${1.12}_{-0.21}^{+0.22}$ & 1.38 & 20 \\ [5pt]
91110-02-16-00 & 0.93 & ${-0.80}_{-0.20}^{+0.14}$ & ${0.16}_{-0.08}^{+0.08}$ & ${1.08}_{-0.27}^{+0.26}$ & 1.39 & 20 \\ [5pt]
91110-02-18-01 & 0.82 & ${-0.70}_{-0.09}^{+0.08}$ & ${0.68}_{-0.12}^{+0.12}$ & ${2.25}_{-0.20}^{+0.20}$ & 1.50 & 20 \\ [5pt]
91110-02-21-00 & 0.81 & ${-0.66}_{-0.05}^{+0.05}$ & ${0.95}_{-0.11}^{+0.10}$ & ${2.70}_{-0.15}^{+0.15}$ & 1.13 & 20 \\ [5pt]
91110-02-41-00 & 0.83 & ${-0.84}_{-0.33}^{+0.19}$ & ${0.18}_{-0.14}^{+0.16}$ & ${1.14}_{-0.44}^{+0.50}$ & 1.09 & 20 \\ [5pt]
92082-01-01-01 & 0.84 & ${-1.00}_{-0.23}^{+0.17}$ & ${0.12}_{-0.07}^{+0.08}$ & ${0.99}_{-0.30}^{+0.35}$ & 0.72 & 20 \\ [5pt]
\midrule
\multicolumn{7}{c}{XTE~J1859$+$226} \\
\midrule
40124-01-48-01 & 0.78 & ${-0.81}_{-0.81}^{+0.34}$ & ${0.14}_{-0.10}^{+0.13}$ & ${1.02}_{-0.37}^{+0.46}$ & 0.73 & 9 \\ [5pt]
40124-01-51-03 & 0.78 & ${-0.99}_{-1.02}^{+0.40}$ & ${0.13}_{-0.09}^{+0.19}$ & ${1.03}_{-0.62}^{+0.76}$ & 0.28 & 9 \\ [5pt]
40124-01-53-01 & 0.83 & ${-1.05}_{-1.41}^{+0.38}$ & ${0.13}_{-0.13}^{+0.31}$ & ${1.05}_{-1.04}^{+1.29}$ & 0.50 & 9 \\ [5pt]
40124-01-53-02 & 0.82 & ${-0.77}_{-0.65}^{+0.32}$ & ${0.22}_{-0.21}^{+0.26}$ & ${1.28}_{-0.60}^{+0.76}$ & 0.36 & 9 \\ [5pt]

\bottomrule
\\
\multicolumn{7}{l}{$Note.$ The parameters are given by model fits to the rms normalized power spectra with the} \\
\multicolumn{7}{l}{model $ P(f) = N f^{index}$.} \\
\multicolumn{7}{l}{$a-$ The fraction of the disk component in relation to the total 3–20 keV unabsorbed flux.}\\
\multicolumn{7}{l}{$b-$ The power-law index}\\
\multicolumn{7}{l}{$c-$ The power-law normalization in the unit of $(rms/mean)^2$.}\\
\end{longtable}
\twocolumn

\end{document}